\documentclass[fleqn,usenatbib]{mnras}

\usepackage{newtxtext,newtxmath}

\usepackage[T1]{fontenc}
\usepackage{ae,aecompl}

\usepackage{graphicx}	
\usepackage{amsmath}	

\usepackage{amssymb}	
\usepackage{xcolor}

\usepackage{etoolbox}

\title[Spectral index of quasar variations]{Dust and the intrinsic spectral index of quasar variations: hints of finite stress at the innermost stable circular orbit}

\author[J. R. Weaver et al.]{
John R. Weaver$^{1, 2, 3}$\thanks{E-mail: john.weaver.astro@gmail.com},
Keith Horne$^{3}$
\\
$^{1}$Cosmic Dawn Center (DAWN)\\
$^{2}$Niels Bohr Institute, University of Copenhagen, Jagtvej 128, 2200 Copenhagen, Denmark\\
$^{3}$SUPA School of Physics \& Astronomy, University of St Andrews, North Haugh, St Andrews, KY16 9SS\\
}

\date{Accepted XXX. Received YYY; in original form ZZZ}

\pubyear{2022}

\begin{document}
\label{firstpage}
\pagerange{\pageref{firstpage}--\pageref{lastpage}}
\maketitle

\begin{abstract}
We present a study of 9\,242 
spectroscopically-confirmed quasars with multi-epoch \textit{ugriz} photometry from the SDSS Southern Survey. By fitting a separable linear model to each quasar's spectral variations, we decompose their five-band spectral energy distributions into variable (disc) and non-variable (host galaxy) components. In modelling the disc spectra, we include attenuation by dust on the line of sight through the host galaxy to its nucleus. We consider five commonly used attenuation laws, and find that the best description is by dust similar to that of the Small Magellanic Cloud, inferring a lack of carbonaceous grains from the relatively weak 2175\,\AA\ absorption feature. We go on to construct a composite spectrum for the quasar variations spanning 700 to 8000\,\AA. By varying the assumed power-law $L_{\nu}\propto\nu^\alpha$ spectral slope, we find a best-fit value $\alpha=0.71\pm0.02$, excluding at high confidence the canonical $L_{\nu}\propto\nu^{1/3}$ prediction for a steady-state accretion disc with a $T\propto r^{-3/4}$ temperature profile. The bluer spectral index of the observed quasar variations instead supports the model of Mummery \& Balbus in which a steeper temperature profile, $T\propto r^{-7/8}$, develops as a result of finite magnetically-induced stress at the innermost stable circular orbit extracting energy and angular momentum from the black hole spin.
\end{abstract}




\section{Introduction}

The optical identification of quasi-stellar objects (quasars hereafter)  by \citet{Matthews1963} 
enabled, for the first time, studies of the distant universe at $z>0.1$. 
Quasars are now recognised as high-luminosity examples of
Active Galactic Nuclei (AGN),
powered by accretion onto a super-massive black hole (SMBH) \citep[][]{1969Natur.223..690L,ShakuraSunyaev1973}.
The continuum variability of quasars, known soon after their discovery, allows us to peer directly into their central engines. Varying by 10-20\% over timescales of months to years, the intrinsic variability of quasar continuum emission has long been theorised to be caused by changes in the environment close to the SMBH. 

Quasar spectral energy distributions (SEDs) provide insight into their underlying physics.
Spanning the full range from gamma rays to radio, quasar SEDs exhibit both thermal (accretion disc, dust) and non-thermal (corona, jet) components.
In the rest-frame UV-optical, thermal emission from the accretion disc is thought to manifest as the `Big Blue Bump' \citep{Shields1978, Malkan1982}, described by a sum of blackbody spectra over a range of temperatures from $\sim10^3$~K for the cool outer edges of the disc to perhaps $\sim10^5$~K near the innermost stable circular orbit (ISCO).  
A related feature is the `Small Blue Bump', caused by closely-packed FeII emission and the Balmer recombination continuum \citep{Wills1985ApJ, Elvis1985}.

For a geometrically-thin  steady-state accretion disc
\citep{ShakuraSunyaev1973}, the effective temperature profile is $T_{\rm eff} \propto \left(M\,\dot{M}\right)^{1/4}\, r^{-3/4}$,
where $M$ is the black hole mass, $\dot{M}$ the accretion rate, and $r$ the radial distance from the black hole.
The corresponding spectrum, obtained by summing blackbody spectra weighted by solid angle, is
$L_\nu \propto
\left( M \, \dot{M} \right)^{2/3}\nu^{1/3}$.
This power-law spectrum applies in the spectral range corresponding to the minimum and maximum disc temperatures, $k\,T_{\rm min} << h\,\nu << k\,T_{\rm max}$,
where $h$ and $k$ are the Planck and Boltzmann constants.
For a more general power-law temperature profile, $T\propto r^{-b}$, the disc spectrum is $L_\nu\propto\nu^{\alpha}$ with $\alpha=(3\,b-2)/b$. Thus measuring the disc's spectral slope $\alpha$  determines the power-law slope $b$ of its temperature profile and tests the accretion disc theory. If the theoretical power-law slope, $\alpha=1/3$ is confirmed, the results measure the product $M\,\dot{M}$.
Moreover, since the disc spectrum scales with inclination angle $i$ and luminosity distance $D_L$ via $\cos{i}/D_L^2$, we may potentially be able to measure quasar luminosity distances\footnote{For Type 1 AGN ($i<60^\circ$),
the mean $\pm$ rms of $\cos{i}$
is $3/4\pm\sqrt{1/48}=0.75\pm0.14$.}.

Several obstacles stand in the way of realising these motivating goals. First, there may be significant extinction and reddening due to dust along the line-of-sight. Correcting for dust in our Milky Way galaxy is relatively straightforward \citep[e.g.,][]{SFD98, SF2011}. More difficult is to correct for the adverse effects of reddening caused by scattering and absorption by dust grains within the host galaxy, a complication shared by the use of Type~Ia supernovae as standard candles. 
Reddening along the line of sight presents a degeneracy since dust grains can redden quasar spectra with a wavelength dependence similar to the power-law form expected for the disc spectrum. However, carbonaceous grains produce an absorption feature prominent in the dust extinction law observed in the Milky Way Galaxy \citep{Allen1976, Seaton1979, Nandy1975} and the Large Magellanic Cloud \citep{Fitzpatrick1986b}. This absorption feature, described by a ``Drude'' profile centered at 2175\,\AA\,\citep{Fitzpatrick1986a, Draine1993}, can largely resolve the degeneracy, given sufficient spectral coverage. The 2175\,\AA\ absorption is weak or absent in other notable systems including the Small Magellanic Cloud \citep{Gordon2003} and local starburst galaxies \citep{Calzetti2000}, which suffer from the full strength of this degeneracy. Moreover, it has been postulated that quasar dust may differ from the varieties studied closely in the local universe,
for example by lacking small grains that are evaporated by the quasar luminosity \citep[e.g.,][]{Gaskell2004}.

Second, the observed spectra of quasars are generally redder than the predicted disc spectrum, hinted at already by \citet{Sandage_1965}. This is due in part to the comparably red starlight of the host galaxy. Quasars are often too distant to directly resolve the host galaxy, which means that their measurements are contaminated by host galaxy starlight captured within the aperture from which the photometry is performed. While this problem can be mitigated in small samples of nearby, resolved quasars where the host galaxy's light profile can be modelled, or extrapolated inward and subtracted from images, this approach fails for larger samples of more distant, unresolved quasars. The advent of large multi-wavelength monitoring campaigns provides a viable workaround. Instead of attempting to subtract host galaxy contributions from imaging data, a time-series of images or spectra can be used to isolate the spectrum of the variable light arising from the central engine and the accretion disc. By this method one can extract separate spectra for the variable accretion disc and the non-varying host galaxy components for a large number of unresolved quasars, provided multi-wavelength photometric monitoring with sub-year cadence over a sufficiently long baseline to probe the variations.

Several successful and innovative campaigns have marked the previous twenty years of the study of AGN. Most recently, assemblages of multi-epoch, multi-wavelength photometric datasets have been observed by the Sloan Digital Sky Survey \citep[SDSS;][]{York2000} and more recently the Zwicky Transient Facility \citep[ZTF;][]{Bellm2018, Graham2019} which have enabled fundamentally new comparisons with theoretical models of accretion disc structure and behaviour with statistically significant samples. Additional observations from the Rubin Observatory's Legacy Survey of Space and Time \citep[LSST;][]{Ivezic2019} will greatly increase both sample sizes and epoch baselines, expected to be underway in 2024. In addition, precise, spectroscopic monitoring campaigns such as the Sloan Reverberation Mapping Project \citep[SDSS-RM; e.g.][]{Shen_2015a} are providing valuable details on the variability of continuum and line emission, although with smaller samples.

Several techniques have been applied to interpret these photometric datasets. For example, \cite{Macleod2012} employed a damped random walk model to describe the stochastic variations for an ensemble of $\sim10^4$ quasars from SDSS, finding good agreement as a viable description of the optical continuum variability. For the same dataset, \citet{Kokubo2014} employed a ``flux-flux correlation'' technique to derive the color of the flux difference spectrum, which was used to infer an accretion disc spectral slope of $L_{\nu} \propto \nu^{1/3}$, consistent with standard steady-state accretion models. Parallel work has been undertaken with this and similar datasets to determine the extinction law most appropriate for quasars \citep[e.g.,][]{Hopkins2004, Krawczyk2015}, the results of which influence work on variability.

The objective of this work is to directly probe the accretion disc light and test theories of accretion physics. This will be accomplished as follows. In Section~\ref{sec:methods} we develop our method that leverages the source variability to isolate the accretion disc light. In Section~\ref{sec:application} we apply this to a sample of 9\,242 quasars observed with multi-epoch multi-wavelength photometry during the Sloan Digital Sky Survey Stripe~82 quasar campaign, including a de-reddening of the isolated accretion spectrum with five commonly used dust extinction laws. 
Section~\ref{sec:composite} presents the composite spectra.
Our results are then discussed in Section~\ref{sec:discussion} and our conclusions made in Section~\ref{sec:conclusions}.

We adopt in this paper a concordance cosmological model with $H_0=70$\,km\,s$^{-1}$\,Mpc$^{-1}$, $\Omega_{\rm M}=0.3$ and $\Omega_\Lambda=0.7$. All magnitudes are in the AB$_\nu$ system \citep{Oke1974}, for which a flux $f_\nu$ in mJy
($10^{-26}$~erg~cm$^{-2}$s$^{-1}$Hz$^{-1}$) corresponds to AB$_\nu=16.4-2.5\,\log_{10}(f_\nu/{\rm mJy})$.

\section{Isolating the accretion disc light}\label{sec:methods}

\begin{figure*}
	\resizebox{\hsize}{!}{\includegraphics[width=\columnwidth]{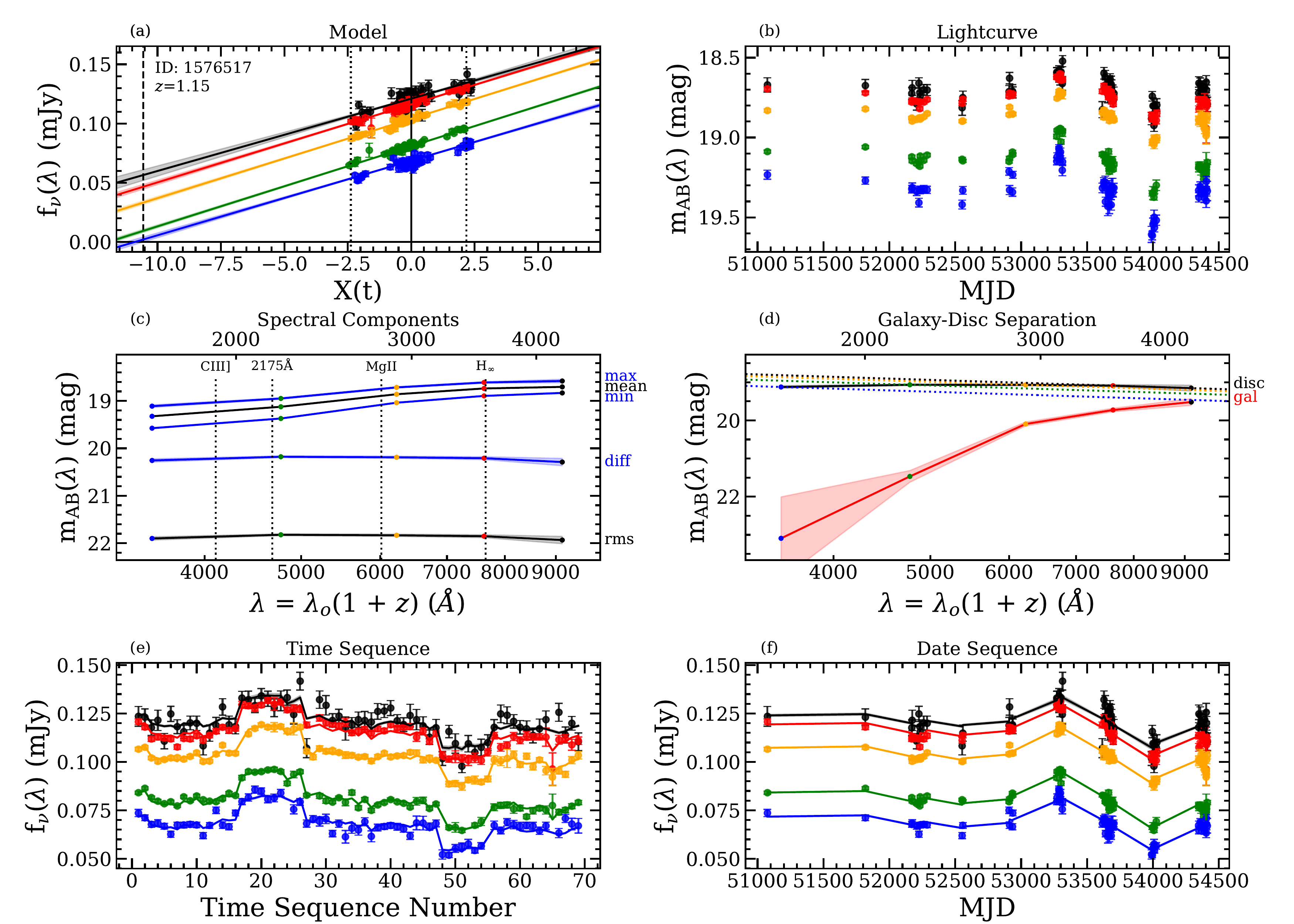}}
	\caption{An example illustrating our lightcurve decomposition method.
	Panel~b shows the $N_{\lambda}$=5-band \textit{ugriz} lightcurve data at $Nt=62$ epochs for the SDSS quasar ID~1576517 at redshift $z=1.15$. Panel~a shows that the flux variations are well fitted by a linear model, $F(\lambda,t) = B(\lambda) + A(\lambda)\, X(t)$, where $B(\lambda)$ is the mean flux, $A(\lambda)$ is the rms amplitude of the flux variations, and $X(t)$ is the lightcurve shape, assumed to be the same for all bands, normalised to $\left<X\right>=0$
	and $\left<X^2\right>=1$.
 The maximum and minimum brightness states are indicated in Panel~a by vertical dotted lines on either side of the mean state at $X=0$.
 Extrapolating to fainter levels, the $u$-band flux becomes negative just below $X\approx-10.5$. Having thus turned off the disc, we attribute the extrapolated fluxes
 at the vertical dashed line, where $u$ is 1-sigma above 0, to the non-variable host galaxy.  Panel~c shows the resulting 5-band spectral energy distributions (SEDs) extracted at the maximum, minimum, rms, and mean states, as well as the bright - faint difference spectrum, with vertical dotted lines marking the wavelengths of
 relevant spectral features.
 Panel~d shows the SEDs extracted for the disc (black) and host galaxy (red). Dotted lines show $L_{\nu} \propto \nu^{1/3}$ fixed to each band and coloured accordingly. Finally, the lightcurve data and fitted model are shown versus time sequence number in Panel~e and by date of observation in Panel~f.  All error envelopes are shown at $\pm1\sigma$.}
	\label{fig:decomposition}
\end{figure*}

In this section we describe  our method using a separable linear model to fit the photometric variations of a quasar observed with multi-wavelength photometry. The method is illustrated for a particular SDSS quasar in Fig.~\ref{fig:decomposition}, which we discuss below as we outline the steps of the analysis.

A quasar is observed at $N_t$ times $t$ in $N_\lambda$ photometric bands, each labelled by its pivot wavelength\footnote{See A2.1 of \citet{Bessell_2012} for details.} $\lambda$.  The observations at time $t$ are considered simultaneous if measured within a time interval so short that changes in the state of the accretion disc can be neglected. For UV and optical observations of quasars this typically means measurements on the same night, or even over a few nights.

We fit the observed spectral flux variations with the following separable linear model:
\begin{equation}
F(\lambda,t)  =  A(\lambda) \, X(t)
+ B(\lambda)
\ .
\label{equ:nband}
\end{equation}
Here $F(\lambda,t)$ can be $F_\nu$ or $F_\lambda$, or indeed any suitable flux unit.
The dimensionless lightcurve shape $X(t)$ is
shifted to zero mean 
and scaled to unit root-mean-square (rms):
\begin{equation}
    \left<\right.X\left.\right>_t=0\ ,
    \hspace{5mm}
     \left<\right.X^2\left.\right>_t=1\ ,  
\end{equation}
where $\left<\right.\cdot\left.\right>_t$ denotes a suitably-weighted time average.
With this normalisation,
the model's amplitude spectrum
 $A(\lambda)$ is the rms of the flux variations about the mean background spectrum $B(\lambda)$. 
 This model has $2\, N_\lambda+N_t$ parameters, which are
 $A(\lambda)$, $B(\lambda)$, and $X(t)$. 
 These are constrained by $N_t \times N_\lambda$ flux measurements plus $2$ normalisation constraints.
 For observations at a single wavelength, $N_\lambda=1$, the model fits the $N_t$ flux measurements exactly.
 For multi-band observations the model parameters are over-constrained by the data, which permits optimising the model parameters by fitting the data, and testing the validity of the model assumptions.

This model fitting is illustrated for a particular SDSS quasar in Fig.~\ref{fig:decomposition}. The lightcurve of this quasar is sampled at $N_{t}=62$ epochs with spectral energy distributions (SEDs) measured by $N_\lambda=5$ bands, as shown in Fig.~\ref{fig:decomposition}b. 
Note here that the data follow a lightcurve shape that is similar for all 5 bands.
Fitting the model to these data, by minimising $\chi^2$, the corresponding set of linear equations is solved to determine the model parameters $A(\lambda)$, $B(\lambda)$, and $X(t)$, with corresponding uncertainties.
This is done in practice by using iterated linear regression fits.
Start by constructing an initial guess for $X(t)$, for example using one of the observed lightcurves, suitably normalised.
Then use 2-parameter linear regression to find $A(\lambda)$ and $B(\lambda)$ assuming $X(t)$ is known. 
Next, revise $X(t)$ assuming $A(\lambda)$ and $B(\lambda)$ are known. Impose the normalisation constraints on $X(t)$. Finally, iterate to convergence.
Some care may be needed to identify and down-weight or reject significant outliers, using a robust procedure such as sigma clipping.

The 2-parameter linear regression fits that determine $A(\lambda)$ and $B(\lambda)$, with $X(t)$ assumed to be known, are presented in Fig.~\ref{fig:decomposition}a. The fitted linear models are shown as solid coloured lines with $\pm1\sigma$ envelopes. The flux data with error bars are plotted versus the dimensionless $X(t)$. This  tracks the changing brightness of the quasar above and below the mean flux level. 
For each band, the slope in this diagram is the rms amplitude $A(\lambda)$
of the flux variations above and below the mean spectrum $B(\lambda)$ at $X(t)=0$.
Note here that the quasar variations are well described by linear flux variations. In particular, there is no evident curvature that could indicate a change in the disc spectrum between the faint and bright states.
The shape of the lightcurve $X(t)$ and comparison of the fitted model with the lightcurve data, are examined in Fig.~\ref{fig:decomposition}e, where the flux data are plotted versus time sequence number, and in Fig.~\ref{fig:decomposition}f versus observation date.
Here the lightcurve shape is determined as a weighted average of the variations seen in all bands.

The fitted model can now be used to predict fluxes expected at different variability states $X(t)$. Given $N_{\lambda}$ observed bands, the fitted model predicts the SED for any variability state $X(t)$. Meaningful extrapolation is possible,
above and below the range sampled by the monitoring data, with the usual caveat that the extrapolated model becomes progressively uncertain. 
Fig.~\ref{fig:decomposition}c
presents the SED obtained for several
indicative states. The mean spectrum, $B(\lambda)$, is the SED derived for the mean state of the system, at $X(t)=0$. 
Above and below the mean SED are SEDs for the faintest and brightest observed states, at $X_{\rm min}$
and $X_{\rm max}$, respectively.
These SEDs are relatively red, $F_\nu$ rising to longer wavelengths.
However, the difference SED between the brightest and faintest states,
evaluated for $\Delta X=X_{\rm max} - X_{\rm min}$,
and the SED of the rms variations, for $\Delta X = 1$, are comparably blue.
Such quasar variations are often described as ``bluer when brighter''. However, the linearity seen in Fig.~\ref{fig:decomposition}a shows that this is not due to the disc spectrum becoming bluer when brighter, but rather to the relatively red (host galaxy) spectrum becoming dominant as light from the relatively blue disc dims.

Our linear model fit to the spectral variations  determines the disc and host galaxy SEDs shown in Fig.~\ref{fig:decomposition}d. The host galaxy SED is obtained by extrapolating the linear model to fainter states until the disc is effectively turned off. 
 Here we define this static point as the variable state $X_{\rm gal}$ at which the lower $1\sigma$ uncertainty envelope of any band is predicted to lie at zero flux. The model would be nonphysical at any fainter state. With short-term variability assumed to arise by modulating the disc's SED, the 
variable disc's SED is the flux emitted at each band in excess of the galaxy's SED.
Note in Fig.~\ref{fig:decomposition}d that the disc SED is close to, but slightly redder than, the power-law $L_\nu \propto \nu^{1/3}$ spectra, indicated by dotted lines.

\section{Application to SDSS data}\label{sec:application}

\subsection{Sample Selection: SDSS Stripe~82 Quasars}

\begin{figure}
	\resizebox{\columnwidth}{!}{\includegraphics[width=\columnwidth]{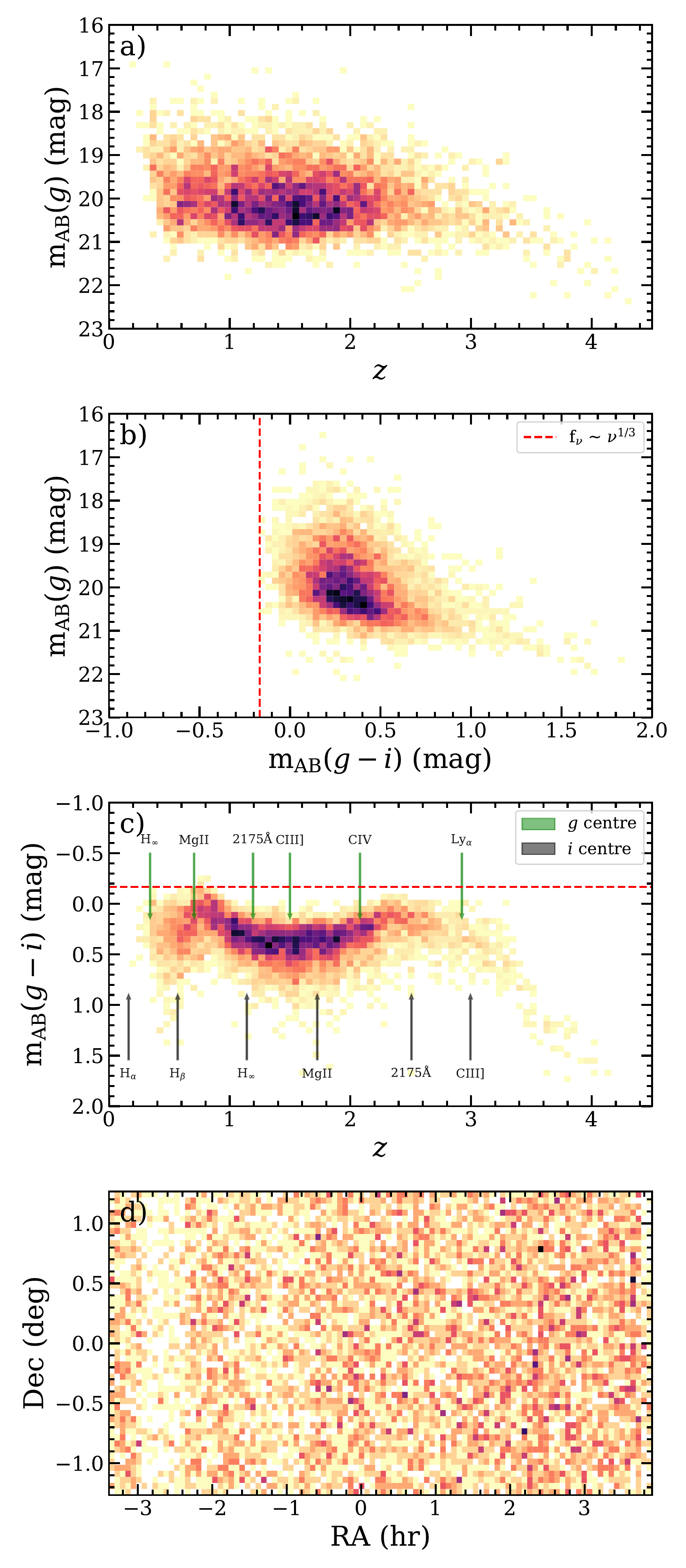}}
	\caption{Summary of the raw input photometry from the SDSS Southern Sample. The $g$-band magnitude and $g-i$ colour distributions with redshift are shown in Panels a and c. The $g$-band magnitude distribution is also shown against the $g-i$ colour distribution. The red dotted line indicates the expected $g-i$ colour from an $\mathrm{f}_{\nu}\,\sim\,\nu^{1/3}$ spectrum. Panel~d shows the density of sources on the sky.}
	\label{fig:sample_selection}
\end{figure}

As described in the previous section and illustrated by Fig.~\ref{fig:decomposition}, our disc + galaxy decomposition procedure requires multi-wavelength coverage with a suitably long time baseline to adequately probe the variability of a given source. More importantly, the procedure is well-posed mathematically if and only if the multi-wavelength coverage is near simultaneous ($\lesssim$ 1 night) as to constrain all relevant regimes of the SED at any one time. Thankfully multi-wavelength photometric coverage for transient surveys are usually performed on nightly basis, thereby providing multiple samplings in wavelength per source, per night. 

A suitable survey satisfying these requirements is the Southern Sample of the Sloan Digital Sky Survey. The Southern Sample catalogue \citep{Macleod2012} contains re-calibrated $ugriz$ lightcurves for all of the spectroscopcially confirmed quasars in SDSS DR7 Stripe 82. Summarily, the catalogue includes 9\,258 quasars over $\sim$290$\,\mathrm{deg}^2$ with an observational baseline of $\sim$10 years, observing each source for $2-3$ consecutive months a year. The total number of epochs per source is $\sim$60 with photometric accuracy between 0.02--0.04~mag. 

The original photometry was adopted from the official SDSS quasar catalogue \citep{Schneider_2010} using PSF magnitudes which were re-calibrated (see \citet{Macleod2012} for details). According to \citet{Schneider_2010}, 97\% of these objects are registered as having point-like morphology, with the remaining 3\% limited to $z\lesssim0.7$; $\sim80\%$ of $z<0.7$ sources are registered as point-like. Future surveys such as LSST will be deeper and have higher angular resolution, relative to SDSS. As such, the task of accurately disentangling the nuclear quasar light from that of resolved host galaxies will require more detailed image modeling and/or aperture photometry.

Fig.~\ref{fig:sample_selection} shows the photometric properties, redshift distribution, and sky density of sources within the catalog. 
This sample provides a broad range in redshift,  $0.1<z<6.0$,  which extends the rest-frame spectral range deep into the ultraviolet and enables us to probe quasar variability and thus accretion disc structure out to remarkably early times.
The photometry here is corrected for Galactic extinction using the coefficients provided in the catalog, to thus be consistent with \citet{Macleod2012}.

As highlighted in Fig.~\ref{fig:sample_selection}b and c, the typical observed-frame $g-i$ colour index of the SDSS quasars, taken from the initial catalogue prior to our decomposition analysis, is $\sim 0.5$~mag redder than a power-law $L_{\nu} \propto \nu^{1/3}$ spectrum. This may be expected due to contamination of the disc spectrum by light from the host galaxy and/or reddening due to dust on the line of sight to the quasar. In addition, the undulating redshift dependence of $g-i$ shown in Fig.~\ref{fig:sample_selection}c likely arises from quasar emission-line features redshifting into and out of the $g$ and $i$ passbands. 

\subsection{Isolating disc SEDs using variations}\label{sec:fitness}

For $99.8\%$ of the Southern Sample, the procedure explained in Section~\ref{sec:methods} succeeded in isolating the variable (accretion disc) and non-variable (host galaxy) SEDs.
 The analysis failed for just $0.2\%$ (16) of the sources, owing to either too sparsely sampled data (either in wavelength or epoch), insufficient variability signal, or a combination of the two.

 The main effect of this disc+galaxy decomposition is evident for Object 1576517 in Fig.~\ref{fig:decomposition}c and d. Much of the red light is ascribed to the non-variable background galaxy component, shown in red in Fig.~\ref{fig:decomposition}d, thus isolating the relatively blue SED of the variable disc light, as shown in black in the same panel. In this case the disc SED is slightly redder than the expected $L_{\nu} \propto \nu^{1/3}$ power-law SED, shown by dotted curves fixed to the observed magnitudes per band. 

\begin{figure*}
$  $	\resizebox{\hsize}{!}{\includegraphics[width=\columnwidth]{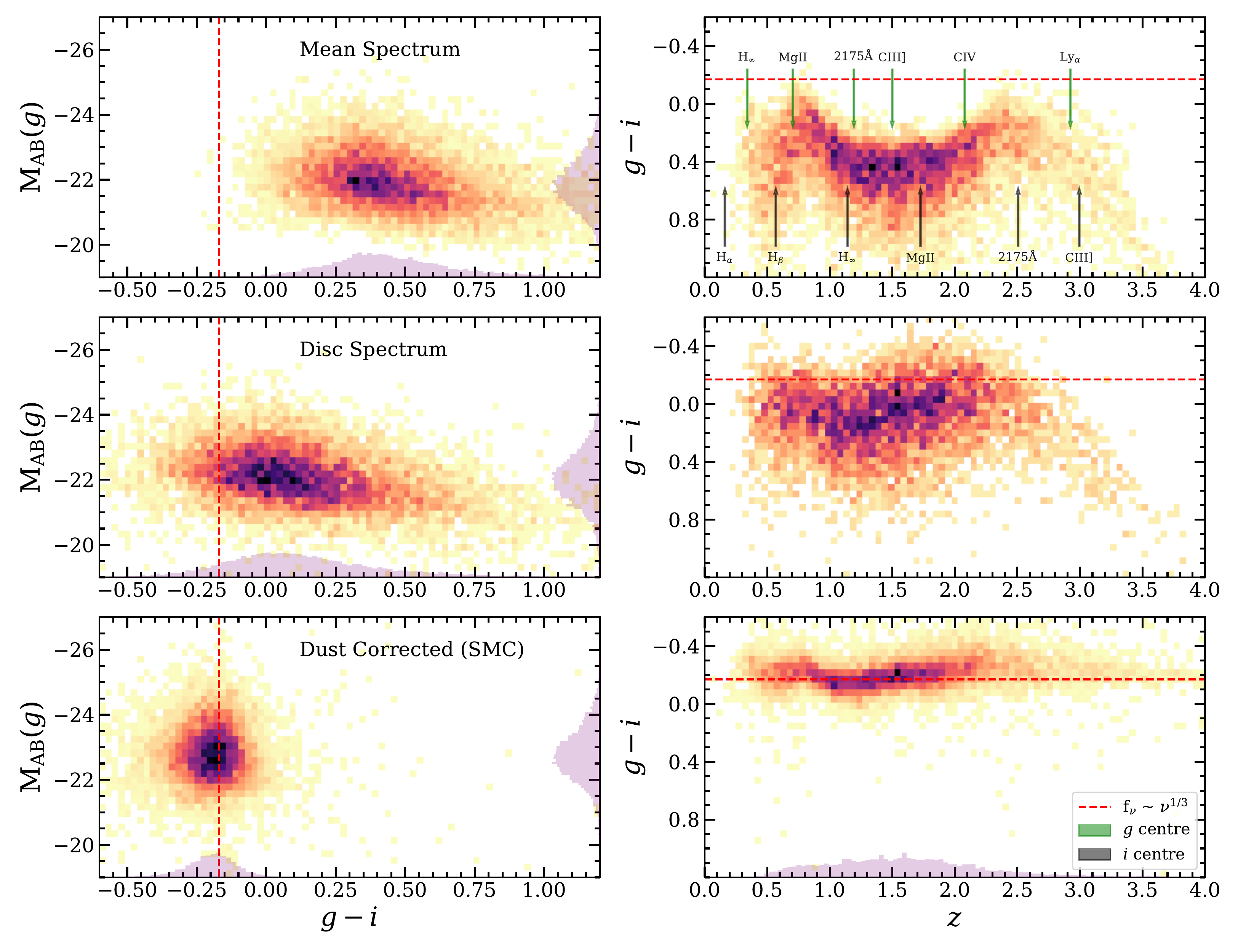}}
	\caption{
	Luminosity (absolute $g$-band AB magnitude) vs $g-i$ color index (left) and  $g-i$ vs redshift (right) for the mean SED(top), disc SED (middle), and \textsc{SMC} de-reddened disc SED (bottom). Distributions are coloured darker with increasing density and shown as histograms projected onto each axis. The $g-i$ colour index for a power-law $L_{\nu}\,\sim\,\nu^{1/3}$ spectrum is indicated by the dashed red lines. Arrows in the top right panel indicate the redshifts at which prominent spectral features are  centered in the $g$- or $i$-band in green and grey, respectively.}
	\label{fig:evol_panels}
\end{figure*}

In Fig.~\ref{fig:evol_panels}, comparing the top and middle panels shows the effect of our disc+galaxy decomposition on $g-i$ colour indices over the full sample of SDSS quasars.
The distribution of $g-i$ colours, for the mean and disc SEDs, are shown here as a function of magnitude and redshift.
A red dashed line marks the $g-i$ colour for the $f_\nu \propto \nu^{1/3}$ power-law.
The $g$-band absolute magnitude distribution is very similar for the disc and mean SEDs, indicating that these quasars are typically brighter at $g$ than their host galaxies.
The $g-i$ distributions differ significantly -- the disc SEDs are generally bluer than the mean SEDs. While none of the SDSS quasars has a mean spectrum as blue as the $L_{\nu} \propto \nu^{1/3}$ power-law, most of the disc SEDs have bluer $g-i$ colours, moving toward and in some cases beyond the $L_{\nu} \propto \nu^{1/3}$ power law colour. But the $g-i$ distribution is not simply translated, rather it appears to be stretched towards bluer colors, leaving behind a long red tail of somewhat fainter quasars with $g-i$ similar in their disc and mean SEDs. One possible interpretation of these redder and fainter SEDs is dust along the line of sight to the quasar disc.
 
Note also that the stark effect of the emission lines causing $g-i$ to undulate with redshift is stronger for the mean than for the disc SEDs.
This is consistent with broad UV emission lines being less variable than the disc continuum.

\subsection{Accounting for dust extinction and reddening}\label{sec:dered}

We now investigate the possibility of dust along the line of sight to the quasar discs. 
This dust could be absent or differ significantly from the dust along lines of sight to other parts of the host galaxy, since the quasar luminosity can heat and evaporate dust in its vicinity. Nevertheless, there is evidence from infrared interferometry  \citep{Honig2013, Asmus2019} 
for both polar dust and equatorial dust. While equatorial dust is thought to obscure the disc and associated broad emission-line regions in Type~2 AGN, polar dust may attenuate and redden the observed disc spectra even for more face-on discs.

Although there is extensive discussion in the literature \citep[e.g.,][]{Gallerani2010, Krawczyk2015, Zafar2015}, there is as yet no definitive evidence and certainly no consensus as to the correct or possibly universal dust extinction law for quasars. Given this uncertainty, we consider 5 possible dust laws. Their attenuation curves, $A_\lambda/A_V$, are shown in Fig.~\ref{fig:dustlaws}.
The 5 dust laws are:

\begin{figure}
	\includegraphics[width=\columnwidth]{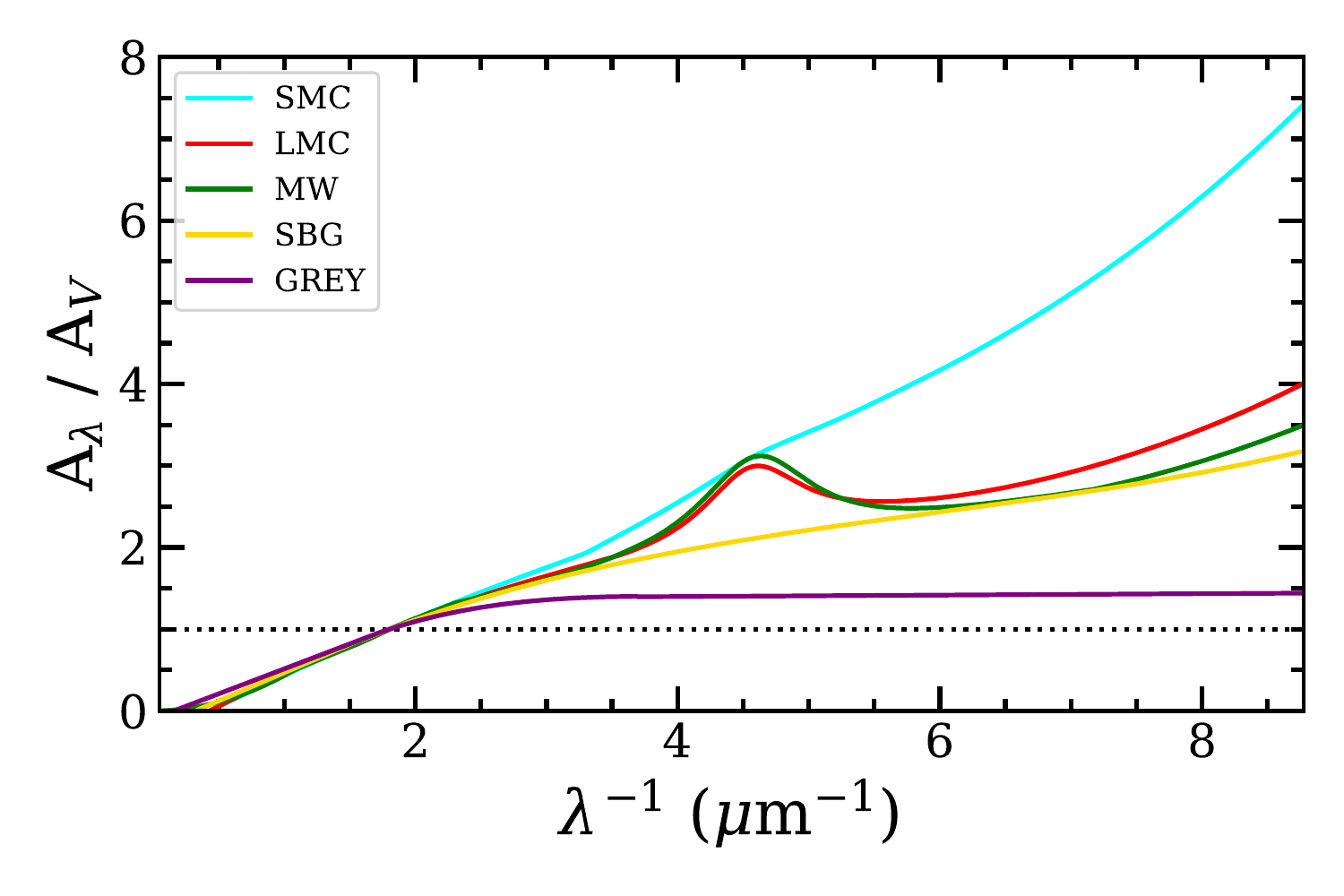}
	\caption{Attenuation curves of five commonly assumed dust attenuation laws used in this work. See \ref{sec:dered} for details.}
	\label{fig:dustlaws}
\end{figure}

\begin{itemize}
\item \textbf{\textsc{SMC}} -- The Small Magellanic Cloud -- a nearly smooth power-law-like curve with relatively high UV extinction due to small grains. Adopted from \cite{Gordon2003}.
\item \textbf{\textsc{LMC}} -- The Large Magellanic Cloud -- a flatter UV extinction curve with a strong 2175\,\AA\, graphite absorption feature. Adopted from \cite{Gordon2003}.
\item \textbf{\textsc{MW}} -- The Milky Way -- Similar to the \textsc{LMC} dust law with a strong 2175\,\AA\,feature. Adopted from \cite{Seaton1979} fitted by \citet{Fitzpatrick1986b}. 
\item \textbf{\textsc{SBG}} -- The Calzetti Starburst Law -- 
a monotonic extinction curve similar to MW and LMC dust but lacking the graphite feature. Adopted from \cite{Calzetti2000}.
\item \textbf{\textsc{GREY}} -- The Gaskell AGN Law -- flattens in the UV due to absence of small grains. Adopted from \cite{Gaskell2004}.

\end{itemize}

The dust-attenuated power-law spectrum model, expressed in absolute AB magnitude vs rest wavelength $\lambda$, is

\begin{equation}
M_{\rm AB}(\lambda)  = 
M_0 + 2.5 \, \alpha \,\log_{10}{\left(\frac{\lambda}{\lambda_0 } \right)} + R(\lambda) \times E(B-V) \ .
\label{equ:dusty}
\end{equation}
Here $R(\lambda)\equiv -2.5 \,\log_{10}{(A_\lambda)}/E(B-V)$ is the dust attenuation in magnitudes per colour excess $E(B-V)$.
The intrinsic power-law spectrum is $L_\nu = L_0\,(\lambda_0/\lambda)^\alpha$, with disc theory predicting a power-law index $\alpha=1/3$.
With no dust, $E(B-V)=0$, the model's absolute AB magnitude is
$M_0$ at the fiducial rest wavelength $\lambda_0=2400$\,\AA,
chosen because the vast majority of the SDSS quasars have rest-frame coverage at $2400$\,\AA\, thus
minimising cases that pivot at wavelengths outside the observed $ugriz$ range.

\begin{figure}
	\resizebox{\hsize}{!}{\includegraphics[width=\columnwidth]{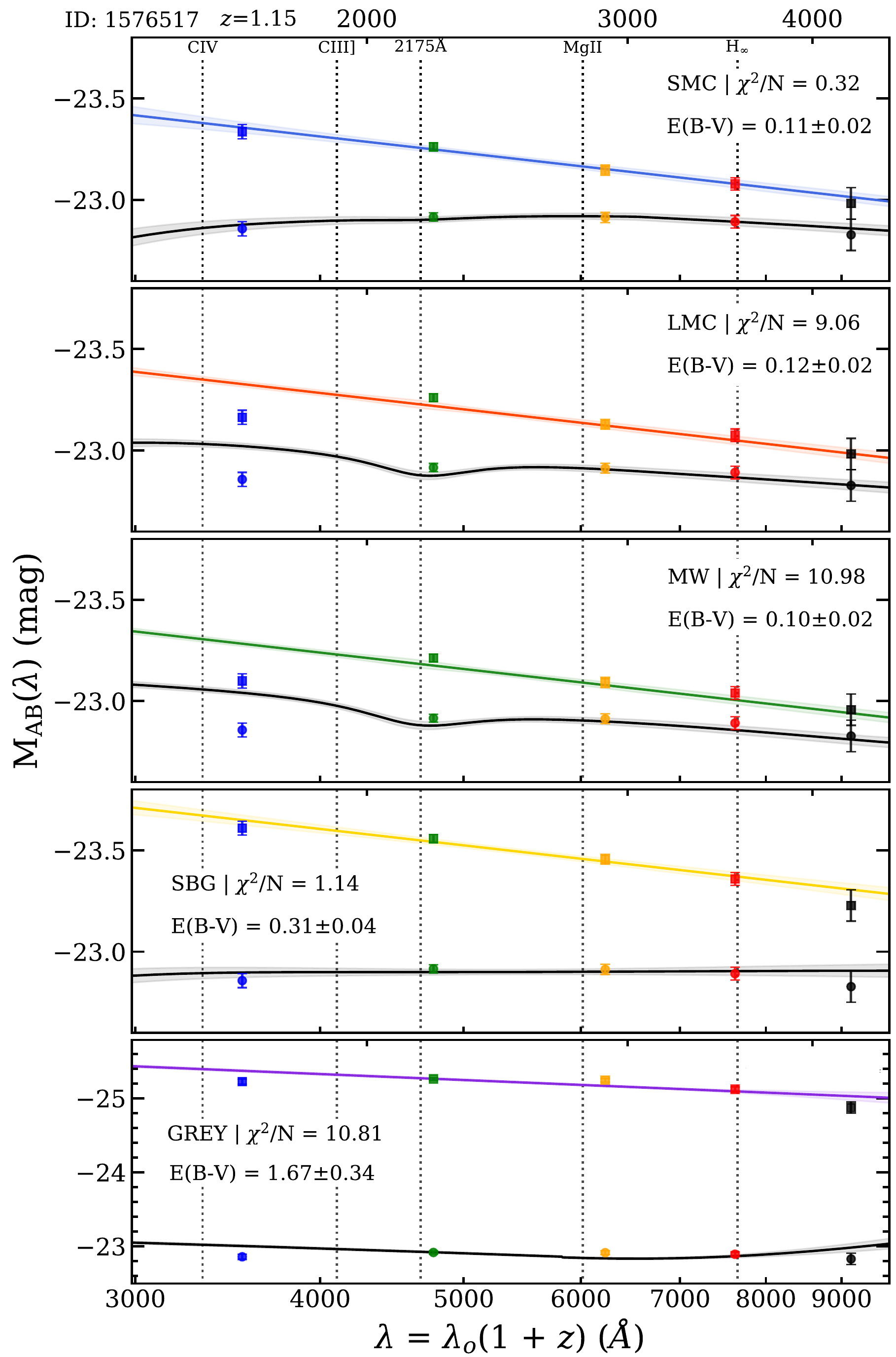}}
	\caption{ Dust-correcting the quasar disc spectrum (absolute AB magnitude versus observed- and rest-frame wavelength) to illustrate the procedure with each of the five dust laws in Fig.~\ref{fig:dustlaws}, for the same object as in Fig.~\ref{fig:decomposition}. In each panel the best-fit dust-attenuated $L_{\nu}\propto\nu^{1/3}$ power-law spectrum (grey curve) is fitted to the observed disc fluxes (filled circles with error bars, coloured to correspond with the $ugriz$ filters). The coloured lines show the same models after dust-correcting by setting $E(B-V)=0$. The corresponding coloured square points are similarly dust-corrected data.
	The best-fit $E(B-V)$ and the reduced $\chi^2/N$ is shown in each panel.
	With 2 parameters fitting 5 data, there are $N=5-2$ degrees of freedom.} 
	\label{fig:dereddening}
\end{figure}

For all 5 dust laws, and for each SDSS quasar, we fit the observed 5-band disc SED, holding $\alpha=1/3$ fixed and minimizing $\chi^{2}$ to estimate the 2 model parameters, $M_0$ and $E(B-V)$ in Eqn.~(\ref{equ:dusty}).
Fig.~\ref{fig:dereddening} 
illustrates this fit and de-reddening procedure
for the disc spectrum of SDSS ID~1576517 ($z=1.15$) determined in Fig.~\ref{fig:decomposition}.
For each of the 5 dust laws, the de-reddened model spectrum, setting $E(B-V)=0$, gives the intrinsic power-law $L_{\nu} \propto \nu^{1/3}$ fixed at the best-fit value of $M_0$. This also allows the photometric data to be dust-corrected by compensating for the dust extinction at each wavelength. This analysis delivers
best-fit estimates for $E(B-V)$ and $M_0$, and a 5-band dust-corrected SED, for each of the 9\,242 quasar discs.

With $2$ parameters fitted to $5$ data, there are $N=3$ residual degrees of freedom. If the data and model are reliable, the reduced $\chi^2/N$ should be $1\pm\sqrt{2/3}$, helping to discriminate among the 5 dust laws. 
For the $z=1.15$ quasar in Fig.~\ref{fig:dereddening},
 the $g$-band happens to sample the redshifted 2175\,\AA\, feature that arises from graphite grains and is prominent in the \textsc{MW} and \textsc{LMC} dust laws.
 The observed disc SED is relatively smooth about the $g$-band. This
 strongly disfavours the \textsc{LMC} and \textsc{MW} dust laws, $\chi^2/N=9.06$ and 10.98 respectively, for which the $g$-band datum is above and the $u$-band datum is well below the best-fit model. 
 For the \textsc{GREY} dust law, the best fit requires a larger dust correction compared with the other dust laws. Also, the \textsc{GREY} dust law leaves relatively large residuals, and so is also strongly disfavoured, $\chi^2/N=10.81$. 
 For this particular quasar, and for the assumed power-law index $\alpha=1/3$,  the \textsc{SMC}
 and \textsc{SBG} dust laws remain viable,
 with $\chi^2/N=0.32$ and 1.14, respectively.

A secondary metric to consider is the best-fit colour excess $E(B-V)$, which quantifies the line-of-sight dust column density. 
A prior on the dust reservoir of the quasar host galaxy may be set by the relatively small values observed in most extragalactic systems, with the notable exception of dusty starbursts \citep{Casey_dsfg, talia_dsfg}. In our analysis, with $E(B-V)$ a free parameter, a fit requiring a much higher $E(B-V)$ should be rightly disfavoured. In Fig.~\ref{fig:dereddening} the  best fit with the \textsc{GREY} dust law gives $E(B-V)=1.67\pm0.34$~mag,
the \textsc{SBG} dust law gives
$0.31\pm0.04$~mag,
 and the \textsc{SMC}, \textsc{LMC}, and \textsc{MW} dust laws are consistent with  $E(B-V)\approx0.11\pm0.02$~mag.
 Thus, importantly, the \textsc{SMC} is not only the best-fit dust-law as measured by $\chi^{2}/N$, it also requires a significantly smaller $E(B-V)$ when compared to the next-best fit \textsc{SBG} dust law.

\begin{figure*}
	\resizebox{\hsize}{!}{\includegraphics[width=\columnwidth]{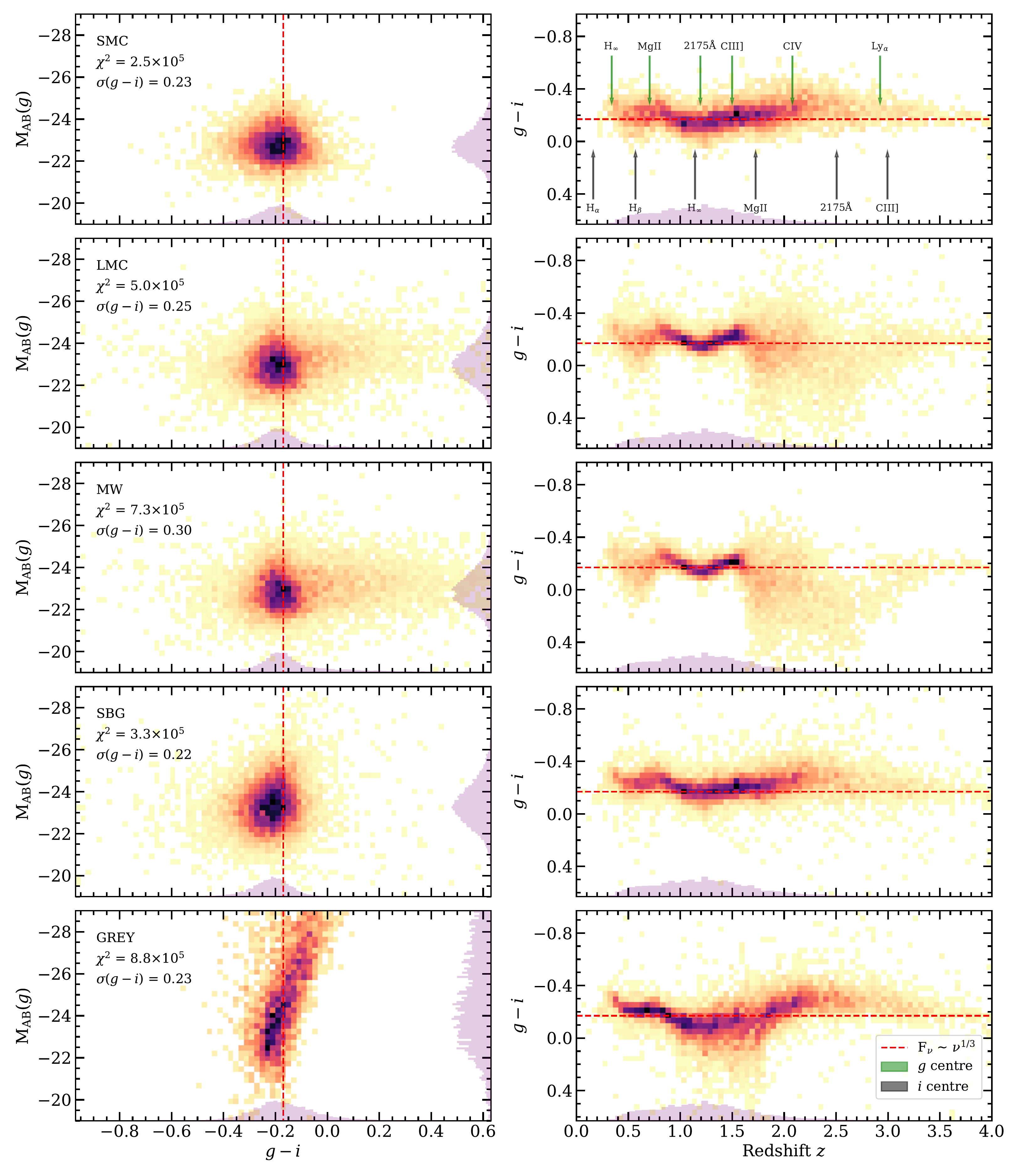}}
	\caption{$g-i$ color-magnitude (left) and redshift-color (right) diagrams for each assumed dust law used to de-redden the 9,156 $ugiz$  disc SED. Distributions are coloured darker with increasing density and shown as a histogram projected onto each axis. The expected $\mathrm{F}_{\nu}\,\sim\,\nu^{1/3}$ is indicated by the dotted red lines. Arrows indicate the presence of an emission feature in the center of the $g$- or $i$-band in green and grey, respectively. The reported values for $\chi^{2}$ are total combined $\chi^{2}$ statistics over the sample.}
	\label{fig:dust_panels}
\end{figure*}

The similarities and differences among the dust law fits discussed above for SDSS ID~1576517 are found to hold statistically in the aggregate sample. 
For the \textsc{SMC} dust law, the lower panels of Fig.~\ref{fig:evol_panels} demonstrate the dramatic tightening of the $g-i$ colour distribution effected by dust-correcting the quasar disc SEDs. 
For all 5 dust laws, Fig.~\ref{fig:dust_panels} compares their dust-corrected colour-magnitude and colour-redshift distributions, reporting for each case the colour dispersion $\sigma(g-i)$ and
the total $\chi^2$ summed over all objects.
The dust-corrected disc SEDs cluster around the assumed intrinsic $L_\nu\propto \nu^{1/3}$ power-law disc spectrum,
with relatively mild dependencies on redshift.
The tightest dispersions, $\sigma(g-i)\sim0.23$ mag,
are achieved similarly by the \textsc{SMC},  \textsc{SBG}, and \textsc{GREY} dust laws.
This is closely followed by the 
\textsc{LMC} and \textsc{MW} dust laws, 
at 0.25 and 0.30~mag, respectively.

Despite their similar success in reducing the $g-i$ dispersion, we note several differences among the 5 dust laws.
First, the \textsc{GREY} dust law, flat in the UV, is problematic as it spreads the dust-corrected disc SEDs over a wide range of implausibly large luminosities.
In our view this strongly disfavours the \textsc{GREY} dust law unless the intrinsic SEDs of quasar discs differ very substantially from a power-law spectrum.

For the \textsc{LMC} and \textsc{MW} dust laws featuring graphite absorption at 2175\,\AA,
the $g-i$ distribution has a tight core 
arising from the redshift range
$0.9<z<1.6$, and broader wings from outside
this range. This redshift structure stems from
the 2175\,\AA\ feature redshifting across the $g$, $r$ and $i$ bands, at $z\sim1.2$, 1.9 and 2.5, respectively.
At these redshifts the evidence for absence of graphite absorption keeps $E(B-V)$ relatively small and better constrained than at intermediate redshifts where the feature falls between bands. This highly structured redshift dependence reduces the viability of our fits with these dust laws (see Appendix~\ref{sec:app1}).

In comparison, the \textsc{SMC} and \textsc{SBG}
dust laws produce dust-corrected disc SEDs with
tight distributions in both luminosity and colour,
with small undulations in redshift that may
plausibly be associated with emission-line features redshifting across the $g$ and $i$ bands, as indicated in the top-right panel of Fig.~\ref{fig:dust_panels}. 
Our fits with these dust laws also achieve the lowest total $\chi^{2}$,
$2.5\times10^5$ for  \textsc{SMC} and $3.3\times10^5$ for \textsc{SBG},
compared with
$(5.0, 7.3, 8.8)\times10^5$
for the (\textsc{LMC}, \textsc{MW} and \textsc{GREY}) dust laws.

In conclusion, the \textsc{SMC} is our preferred dust law. It appears to be both reasonable and the best-fit dust law overall, with a tight $g-i$ color distribution centered about colour of an expected $L_{\nu} \propto \nu^{1/3}$ power-law which is well constrained nearly equally at all redshifts. 
The \textsc{SBG} dust law is a close second choice, but with a somewhat higher $\chi^2$. For individual sources, the SMC provides the best-fit in $43\%$ of cases, followed by the LMC, MW, and SBG at around $\sim17\%$ each, and lastly by GREY at $<7\%$ (see Fig.~\ref{fig:dustfractions}).  We continue with all 5 dust laws, but consider the \textsc{SMC} dust law to be the most appropriate for our subsequent analysis.

\begin{figure*}
	\resizebox{\hsize}{!}{\includegraphics[width=\columnwidth]{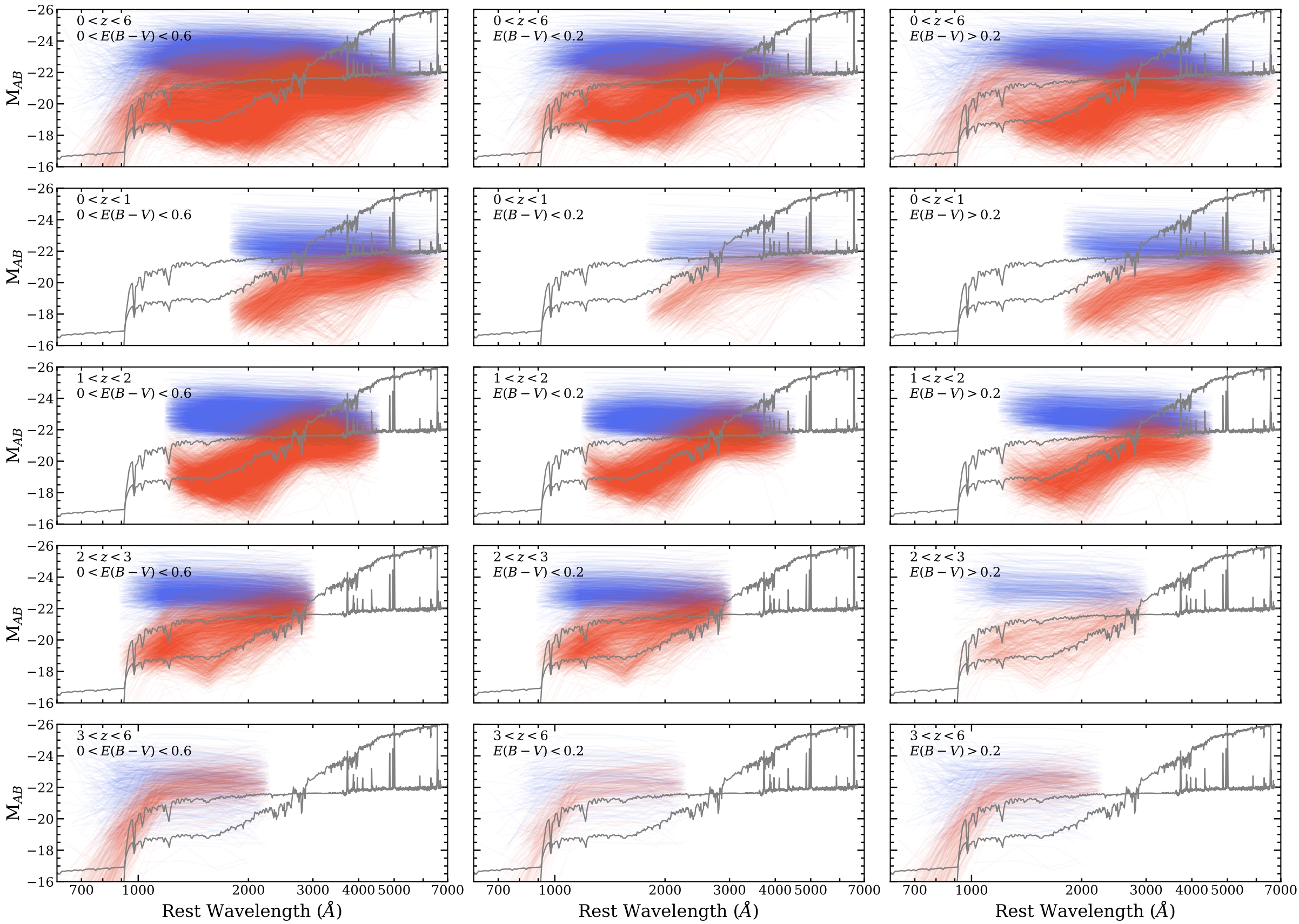}}

	\caption{ Quasar host galaxy SEDs (red) and dust-corrected disc SEDs (blue)
 sorted by redshift (columns) and $E(B-V)$ (rows).
 SEDs for typical galaxies in the red sequence (e.g., the lenticular shell galaxy NGC~7585)
 and blue cloud (e.g., the compact starburst Mrk~930), adopted from \citet{Brown2014}, are shown in each panel for reference.
 }
	\label{fig:galsed}
\end{figure*}

\begin{figure}
	\resizebox{\hsize}{!}{\includegraphics[width=\columnwidth]{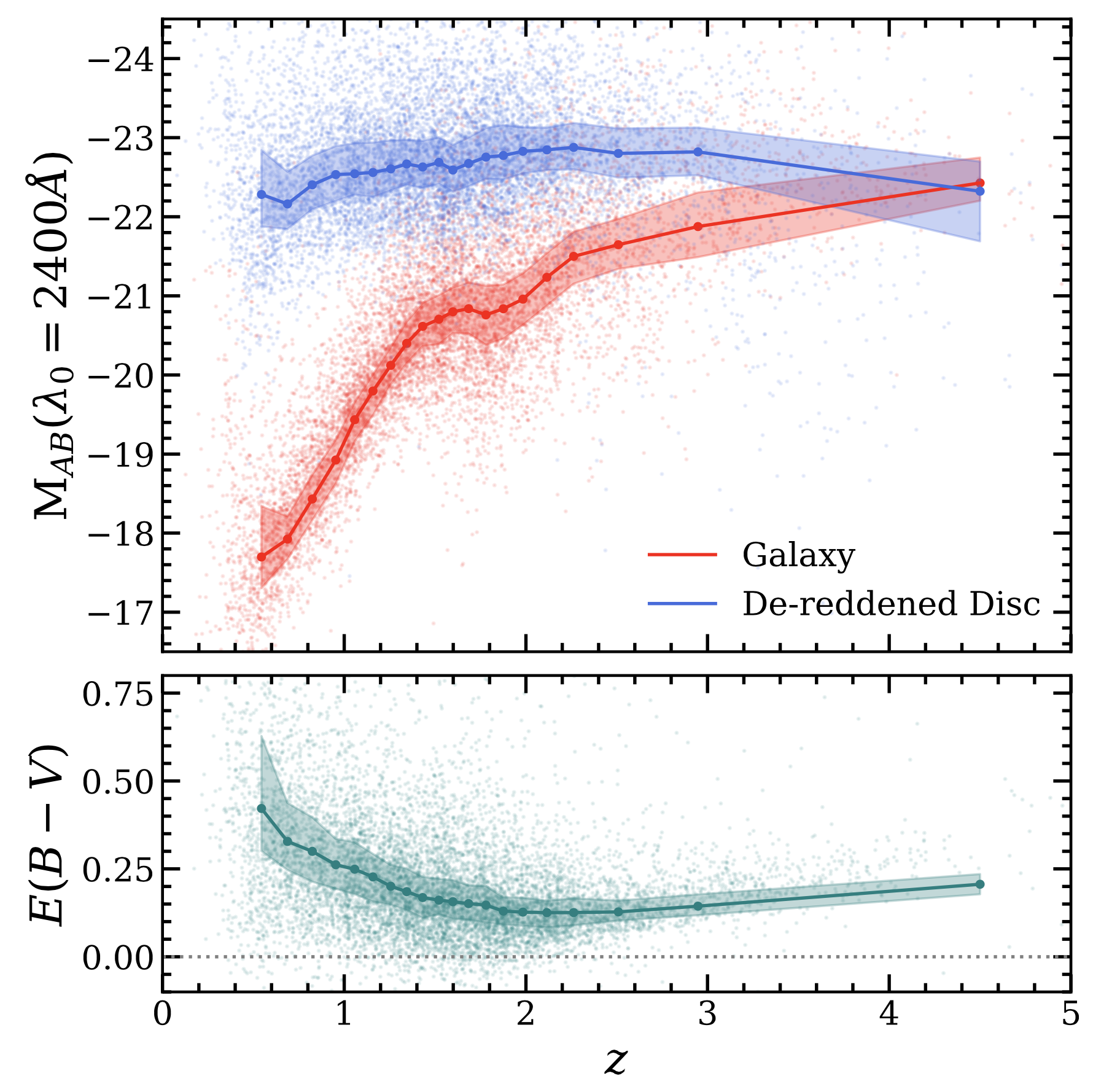}}

	\caption{\textit{Upper:} Rest-frame 2400\,\AA~absolute AB magnitudes as a function of $z$ for the galaxy SED component (red) and disc component (blue), dust-corrected assuming an SMC-like attenuation and $L_\nu\propto\nu^{1/3}$ power-law. \textit{Lower:} Dust attenuation $E(B-V)$ estimates assuming SMC. Bins have equal occupation at all redshifts and are used to determine the medians and 68\% ranges as shown.
 }
	\label{fig:galamag}
\end{figure}

\subsection{ Host Galaxy SEDs }
\label{sec:host}

As a check on our SED decomposition procedure, using variability to separate the variable disc and non-variable galaxy SEDs, we examine the resulting galaxy SEDs.
If our linear extrapolation to (sometimes much) lower fluxes than
observed is a poor approximation, 
the resulting galaxy SEDs could be distorted.

Fig.~\ref{fig:galsed}
shows the SEDs inferred for the quasar host galaxies, sorted by redshift and by dust extinction. The red curves show the galaxy SEDs.
The blue curves show the corresponding dust-corrected disc SEDs.
Higher redshift host galaxies appear to be more luminous than those at lower redshifts. This is a natural consequence of the SDSS quasar sample being approximately magnitude limited, with fainter objects being detectable only at lower redshifts.
  Note that at the highest-redshifts, $3<z<6$, the galaxy SEDs are strongly affected by the Lyman break moving into and thus suppressing the luminosity in the $u$ band.

 The host galaxy SEDs may be expected to be fainter and redder in quasars for which a large $E(B-V)$ is inferred to produce a $L_\nu\propto\nu^{1/3}$ intrinsic disc spectrum.
 However, comparing the right two columns in Fig.~\ref{fig:galsed},
we see no strong trend in this direction. The host galaxies of more attenuated discs are perhaps a bit fainter, but not much redder.
This implies that dust along the line of sight to the quasar disc is not strongly correlated with dust on lines of sight to stars in the host galaxy.

Turning to trends with redshift,
at $z < 1$, the quasar host galaxy SEDs covering 2000 to 6000\,\AA\ all look similar.
They are fainter than and intermediate in spectral shape between the red SED of NGC~7585
and the blue SED of Mrk~930, typical red sequence and blue cloud galaxies,
shown for comparison in each panel of Fig.~\ref{fig:galsed}.
 At $z > 1$, the quasar host galaxy SEDs are brighter, and an increasing fraction of them exhibit a UV component producing a V-shaped $ugr$ dip, with $g$ fainter than $u$ or $r$.
 This can be interpreted as a young stellar population as in star-forming (blue cloud)
or intermediate (green valley) galaxies.
 They constitute a minority at $1 < z < 2$, and a majority at $2 < z < 3$, compatible with maximum star formation at cosmic noon,
and decreasing thereafter. 
At $z>3$ virtually all of the quasar host galaxies
are blue-cloud starbursts, with strong UV emission and brighter than the SED of the compact blue starburst galaxy Mrk~930. The Lyman break appears to depress the galaxy SEDs on the blue end.

These trends with redshift accord with our current understanding of the star formation history of
galaxies over cosmic time \citep{madau_araa, Schreiber_aara}. As summarized in Fig.~\ref{fig:galamag}, star-forming hosts become increasingly faint with time. We find no significant change if we remove the 3\% of sources registered with resolved morphologies, as they constitute $\lesssim20\%$ of sources at $z\lesssim0.7$, and $\ll1\%$ at higher redshifts. While these trends could be affected by unknown selection biases, they are broadly consistent with the well-known fading of star formation between $z\sim2$ and the present epoch.
In contrast to the quasar host galaxies, the dust-corrected disc luminosities are remarkably stable across all epochs, $M_{\rm AB}\sim-22.6$ at $\lambda_0=2400$\,\AA,
with a dispersion of $\sim0.4$~mag, becoming
less certain at $z>3$ where extrapolation redward of the observed SED is required. 
Nevertheless, these encouraging results serve to validate our procedure using variability to separate the quasar disc and host galaxy light.

 Some of our galaxy SEDs have $u$ brighter than $g$, in fact rising
more rapidly into the UV compared
 with a blue stellar population, perhaps even more rapidly than a Rayleigh-Jeans slope. This effect is likely a small flaw in our decomposition procedure.
 We currently set $X_{\rm gal}$ at the lowest possible level, so that the extrapolated
flux in one band, usually $u$ or $g$, is 1-$\sigma$ above 0. 
 A slightly higher level for $X_{\rm gal}$ could be used, thus moving a small fraction of the disc SED to the galaxy SED. The effect would be to make the
V-shaped $ugr$ dip in the galaxy SEDs less prominent in those cases where $g$ is fainter than $u$, elevating the galaxy flux at $g$ and moving the galaxy SEDs closer to the SED of a blue cloud galaxy. The disc SED would then have a correspondingly lower flux at $g$. 
We have not yet implemented this procedural tweak. We expect it to have a relatively small effect on the disc SEDs, which are much brighter than the galaxy.

These results follow the trends found in the analysis of \citet{Matsuoka2014} who performed a spatial decomposition to extract point-like quasar signals from their host galaxies, based on the same SDSS observations of Stripe 82. Limited to resolved sources at $z<0.6$, they find that quasars are bluer than their host galaxies, with a quasar-to-host ratio of $\sim8$ in $u$ and $\sim1$ in $i$. For our sources at $z<0.6$, host light is also typically fainter than our de-reddened discs, by a factor of $\sim90$ in $u$ and $\sim3$ in $i$. However, a more equivalent comparison using our reddened (i.e. uncorrected) disc components produces a less extreme ratio of $\sim30$ in $u$ and $\sim2$ in $i$, in better agreement with \citeauthor{Matsuoka2014}. The remaining discrepancy could be driven by a combination of selection effects, PSF-modelling biases in the analysis from \citeauthor{Matsuoka2014}, and that our definition of host galaxy may underestimate the host contribution at $u$. Regardless, it seems that the quasar discs are bluer than their hosts.

\section{Composite Spectra of Variable Quasar Discs}\label{sec:composite}

The SDSS Stripe~82 quasar sample provides an unprecedented multi-year record of
multi-band quasar variations, but it is limited to just five optical bands ($ugriz$). Despite this drawback, we can leverage the cosmological redshift range to construct a composite quasar disc spectrum at somewhat finer spectral resolution and extending to much bluer rest-frame ultraviolet wavelengths. Our approach implicitly assumes that the spectral features of the accretion disc are universal, an assumption we made in the dust-correction procedure by assuming an $L_\nu \propto \nu^{1/3}$ power law for the intrinsic disc spectrum. As justified in Section~\ref{sec:dered}, we assume that local extinction follows the \textsc{SMC} law for all sources when constructing our composite spectrum. It is also important to note that spectral features seen here will be smoothed out by the resolution of the filter profile of each band. 

For each SDSS quasar, we have removed the host galaxy contribution by using the spectrum of the variable component, and corrected for possible dust extinction and SMC-like reddening in the host galaxy, assuming an $L_\nu \propto \nu^{1/3}$ power-law for the intrinsic disc spectrum.
We construct a composite disc spectrum by combining the resulting disc SEDs for quasars sampled across a continuous range of redshifts $0.1<z<6$.
Our dust-correction procedure fits the power-law model, $L_\nu(\lambda) = L_0\, \left(\lambda_0/\lambda\right)^\alpha$,
assuming $\alpha=1/3$, to determine for each quasar
the luminosity $L_0$
at reference rest wavelength $\lambda_0=2400$\,\AA,
and the required $E(B-V)$.
The power-law spectrum provides a backbone for our composite disc model.
We simply scale the $L_0$ for each quasar to a common value (-22.6\,AB, as evidenced by Fig.~\ref{fig:galamag}), and scale the dust-corrected $ugriz$ fluxes by the same
scale factor. 
This provides five measurements on the composite spectrum, at the rest wavelengths of the $ugriz$ bands at the redshift of that quasar.
Doing that for the full sample gives $\sim45\,000$ points, which we then average with a binned median to reduce the scatter.

In Section~\ref{sec:canonical}, we construct the composite disc spectrum assuming the canonical $L_\nu\propto\nu^{1/3}$ power-law. 
In Section~\ref{sec:alpha} we generalise this analysis by assuming $L_\nu \propto \nu^{\alpha}$ and solving for the best-fit power-law slope $\alpha$, thus testing the disc theory prediction $\alpha=1/3$.

\subsection{ Composite disc Spectrum for $L_\nu \propto \nu^{1/3}$ }
\label{sec:canonical}

\begin{figure*}
	\resizebox{\hsize}{!}{\includegraphics[width=\columnwidth]{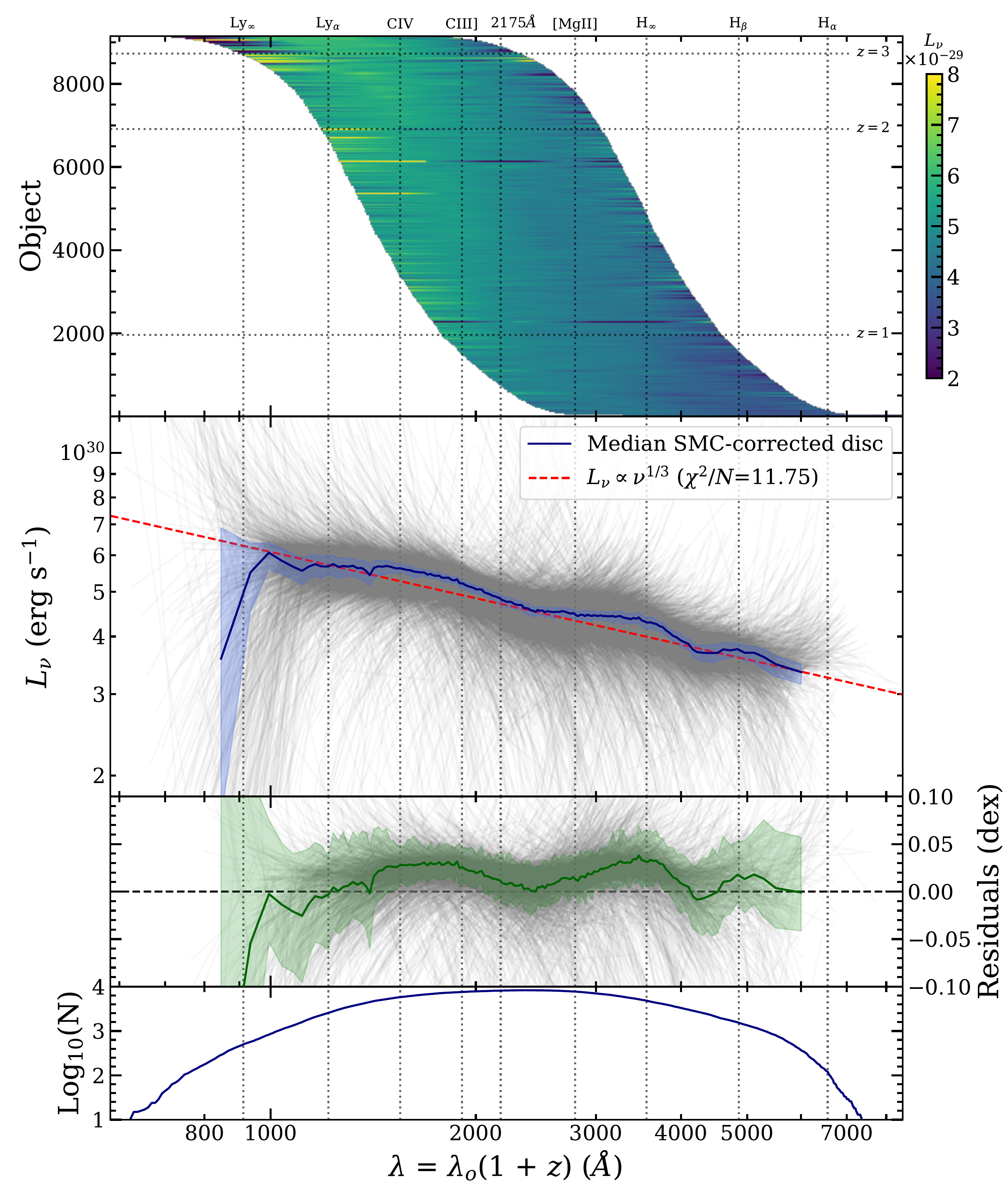}}
	\caption{Spectral energy distributions for the sample are shown at the top, shifted into the restframe and ordered by $z$. Relevant emission lines are indicated by the vertical dotted lines. Shown in the middle is the composite spectrum of the de-reddened disc component computed with a binned median (blue, $\sim300$ points per bin) with an 68 percentile envelope indicating the width of the distribution at that point, with scatter indicated behind by 1-in-5 SEDs. We assume an \textsc{SMC}-like reddening and $L_{\nu}\propto\nu^{1/3}$ power law, overlaid in red. The residuals are shown in  the lower panel in green, with an envelope likewise from above. The lower panel shows the distribution of sources contributing to any give rest-frame wavelength.}
	\label{fig:composite_spectrum}
\end{figure*}

Fig.~\ref{fig:composite_spectrum} presents the composite disc spectrum assuming a canonical $L_\nu\propto\nu^{1/3}$ power-law. We remove 1\% of sources with the worst $\chi^2$ (typically in excess of 100) to keep these poor fits from dominating the overall statistics. The top panel stacks the 5-band dust-corrected disc SEDs for over the remaining 9\,150 SDSS quasars, sorted by redshift, and coloured according to relative brightness, interpolating linearly between the pivot wavelengths of the $ugriz$ bands. Despite somewhat larger noise in the $u$ and $z$ bands, there is clearly a general increase in brightness toward bluer restframe wavelengths. This reflects the assumed $L_{\nu} \propto \nu^{1/3}$ power-law adopted for the dust corrections. 

The middle panel presents the composite disc spectrum, which
undulates above and below the assumed power-law spectrum (red dashed line).
Here the $5\times9\,150$ individual photometric measurements are summarised using a median with $\sim300$ points in each bin (blue curve with an 68\% envelope). We additionally show 1 in 5 (i.e., 20\%) of sources to illustrate the object-to-object variations.
The panel below this uses a similar format (in green) to show residuals relative to the power-law.
The bottom panel shows the number of quasars contributing at each wavelength, nearly all 9\,150 in the middle at 2400\,\AA\ and dropping below 100 on the ends below 600\,\AA\ and above 7000\,\AA. 

 The power-law model provides a reasonable match to the data, with a reduced $\chi^{2}/{N}=11.75$. 
Undulations
around the power-law are significant and plausibly attributed to variable spectral features such as
the Balmer continuum emission around 3500\,\AA. 
Only a handful of low-redshift quasars contribute to the H$\alpha$ peak at 6563\,\AA.
The downturn blueward of 1200\,\AA\
is expected due to intervening Lyman $\alpha$ forest
absorption in the $u$-band at $z>2$. 
The more dramatic drop blueward of 900\,\AA\ 
is the Lyman break arising from Lyman continuum absorption depressing the $u$-band flux in the highest-redshift quasars at $z>3$.

\subsection{Consideration of alternative spectral slopes}
\label{sec:alpha}

\begin{figure*}
	\resizebox{\hsize}{!}{\includegraphics[width=\columnwidth]{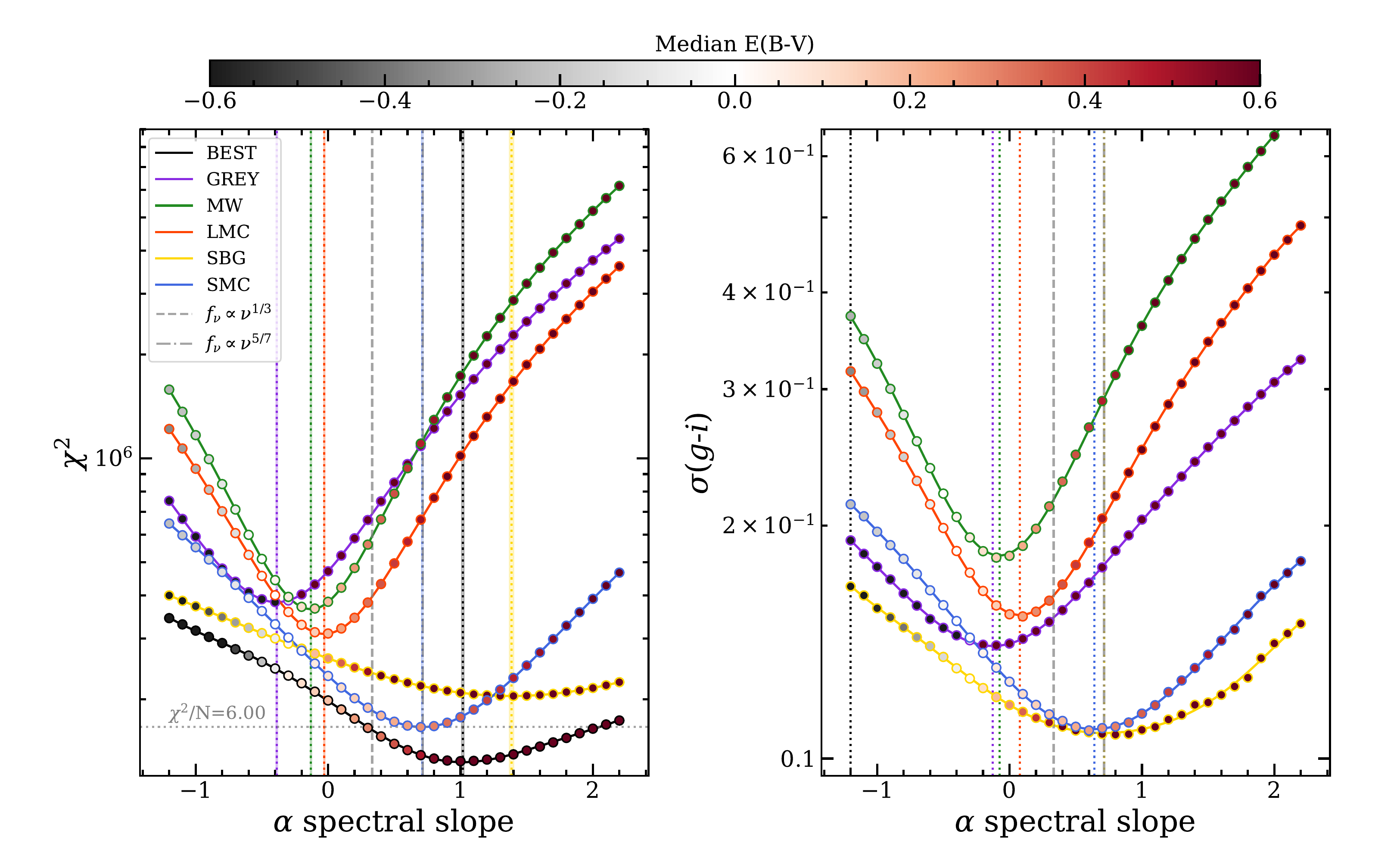}}
	\caption{Two aggregate badness-of-fit (BoF) metrics, the total $\chi^{2}$  (left) and $\sigma(g-i)$ (right), as functions of the accretion disc spectral index $\alpha$. Points show the BoF values for each of 5 candidate dust laws, fitting the $\alpha$ grid with 7$^{th}$ order polynomials.
	 For each dust law the median $E(B-V)$ is indicated by the filler of each point, red for positive values and grey for (nonphysical) negative attenuation. The best-fit $\alpha$ values, at BoF minima, are marked on the left panel by vertical dotted coloured lines within coloured bands denoting
	the uncertainty in $\alpha$ based on the $\Delta\chi^{2}=\chi_{\rm min}^{2}/{N}$ criterion.
	 Fiducial power-law models corresponding to $\nu^{1/3}$ and $\nu^{5/7}$ are marked by vertical  grey lines. The best fit is achieved for SMC-like dust, median $E(B-V)=0.28$ and $\alpha=0.71\pm0.02$, close to $5/7$, as detailed in Table~\ref{tab:bestfit_params}. Lastly we compute the total $\chi^2$ at each $\alpha$ using value corresponding to the best-fit dust law for each source, shown in black circles.}
	\label{fig:bestfit_chisq}
\end{figure*}

\begin{figure}
	\resizebox{\hsize}{!}{\includegraphics[width=\columnwidth]{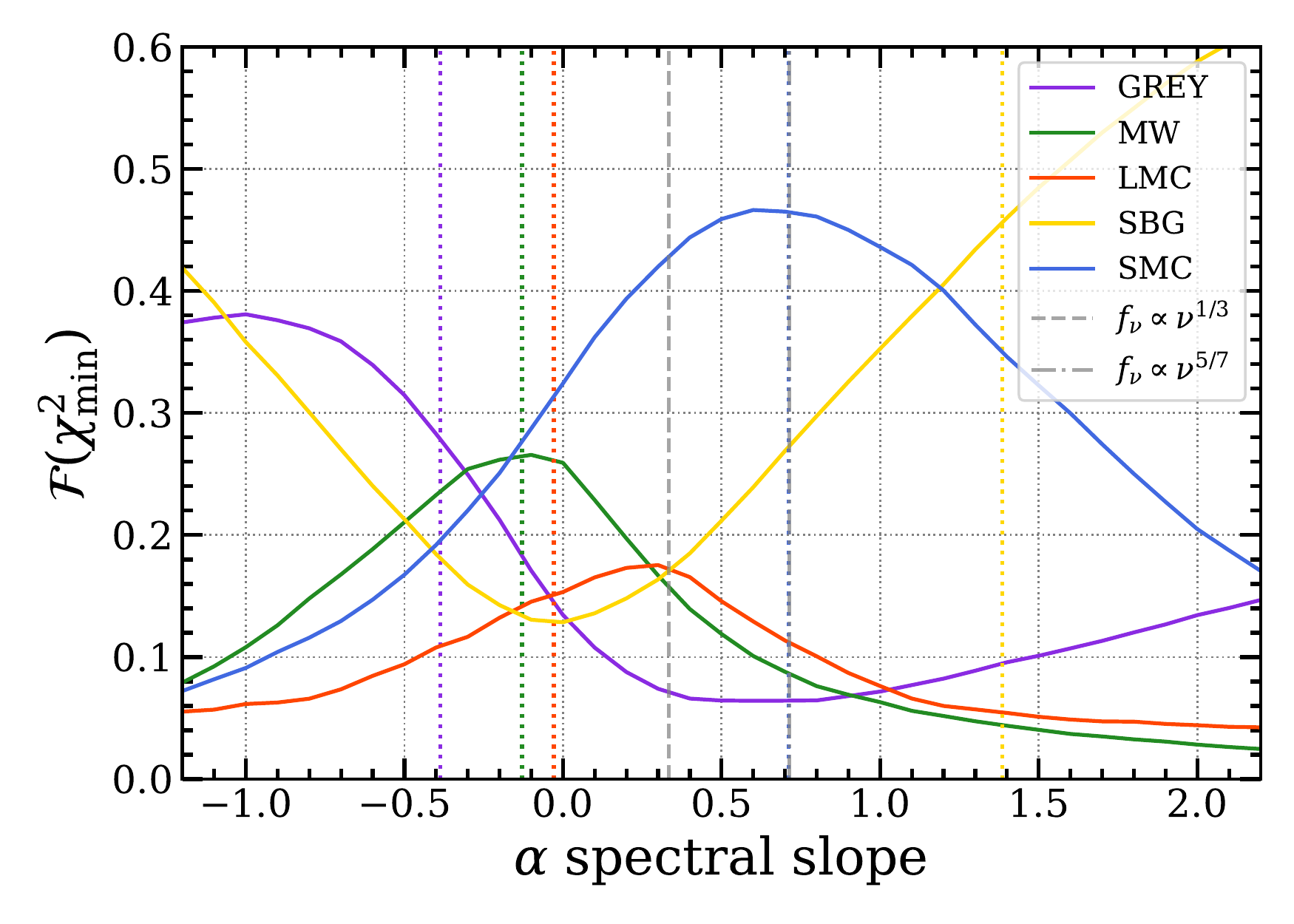}}
	\caption{Fraction of sources per assumed spectral slope $\alpha$ best-fit by the given attenuation law, as measured by minimum $\chi^{2}$. Vertical dotted lines indicate the best-fit $\alpha$ for each dust law.} 
	\label{fig:dustfractions}
\end{figure}

\begin{figure*}
	\resizebox{\hsize}{!}{\includegraphics[width=\columnwidth]{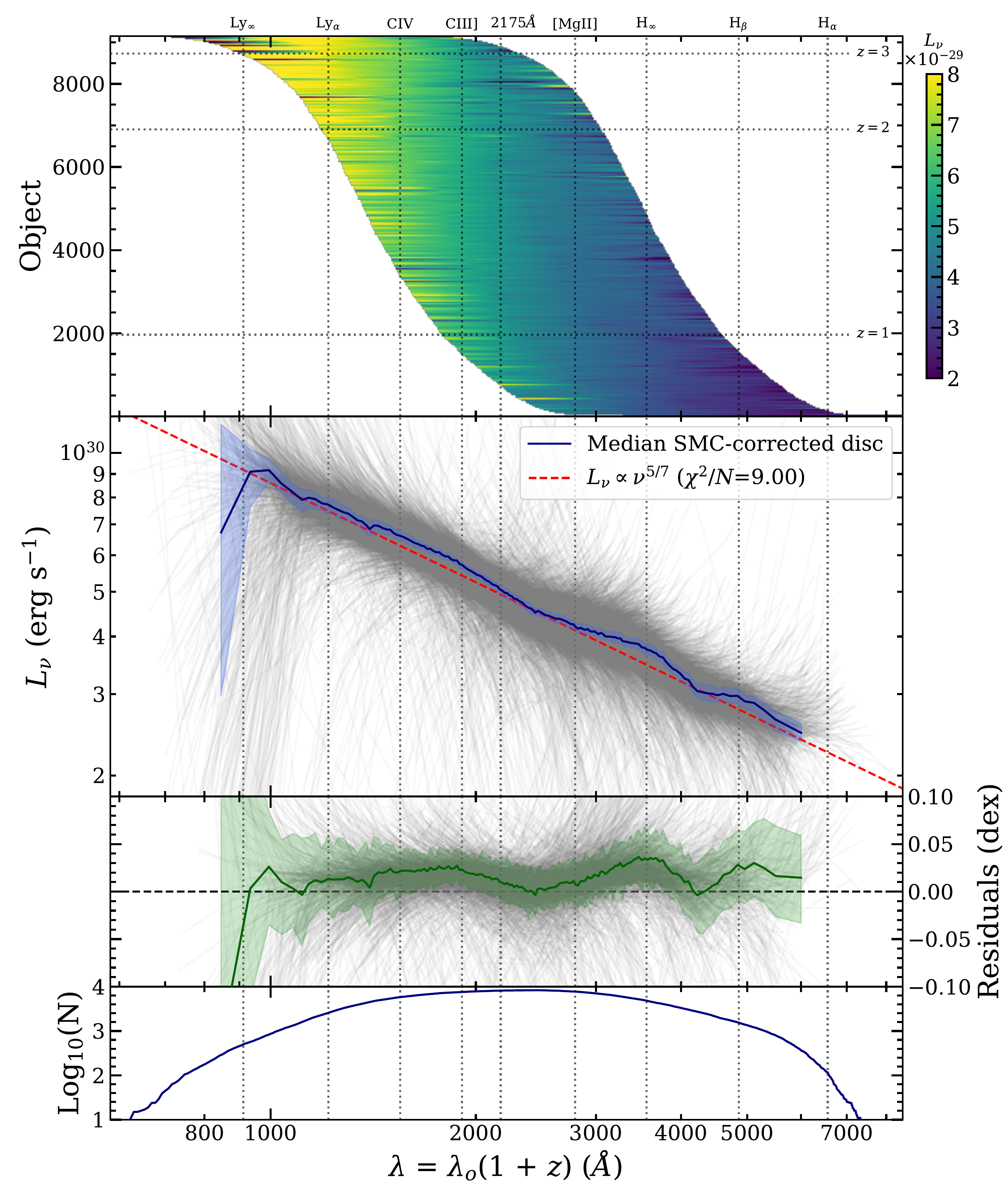}}
	\caption{Same as Fig.~\ref{fig:composite_spectrum}, but assuming a $L_{\nu} \propto \nu^{5/7}$ powerlaw.}
	\label{fig:composite_spectrum2}
\end{figure*}

The common assumption that the underlying powerlaw should be $L_{\nu} \propto \nu^{1/3}$ originates from classical theory \citep{ShakuraSunyaev1973} and has not been conclusively demonstrated to be the true underlying spectral power-law. This work so far has made the same assumption, and so now we question it. While the disc-decomposition procedure is entirely independent of the assumed underlying power-law, the de-reddening procedure is not. Hence, we re-derive all dust reddening solutions for each of the five assumed dust laws assuming a range of underlying power-laws slopes.

As before, we compute the aggregate $\chi^{2}$ statistic for each assumed dust law \textit{and} underlying power-law index $\alpha$ such that $L_{\nu} \propto \nu^{\alpha}$, for a coarse grid of $\alpha$ values. The results are shown in Fig.~\ref{fig:bestfit_chisq}, measured both using the aggregate $\chi^{2}$ of the sample (left) and only the scatter in $g - i$ colours (right). Each coloured curve corresponds to an assumed dust law and each point in $\alpha$ is coloured by the median $E(B-V)$ achieved under those two assumptions.  Whereas red colours indicate high median dust extinction, greyscale colours denote the median $E(B-V) < 0$ is non-physical, although some objects in a given collection may still have $E(B-V) > 0$. The best-fit $\alpha$ for each curve is calculated by fitting a 7\textsuperscript{th} order polynomial to the samples in $\alpha$ and computing the minimum $\chi^{2}$. Estimates for the best-fit values of $\alpha$ are shown also in Fig.~\ref{fig:bestfit_chisq} and are indicated by color-corresponding vertical dotted lines with their uncertainty envelopes calculated where $\Delta\chi^{2}=\chi_{\rm min}^{2}/{N}$, which are unexpectedly small for such a large data set. For reference, grey vertical dashed and dash-dotted lines are included to indicate $\alpha$ corresponding to the canonical $L_{\nu} \propto \nu^{1/3}$ as well as an alternative $L_{\nu} \propto \nu^{5/7}$, respectively.
Table~\ref{tab:bestfit_params} summarises the best-fit parameters for each dust law,
for fits with $\alpha=1/3$, $\alpha=5/7$, and $\alpha$ optimised for each dust law.

\begin{table} 
\setlength{\tabcolsep}{5pt} 
\begin{tabular*}{\columnwidth}{l|ccccc}
\hline\hline
& $\alpha$ & $\chi^{2}/{N}$ & $E(B-V)$ & $E(B-V)$ & $\sigma(g-i)$ \\
& & & Median & $\sigma_{\rm MAD}$ &  \\
\hline
&\multicolumn{4}{l}{$L_{\nu} \propto \nu^{1/3}$} \\
\hline
SMC & $0.33\ldots$ & 6.70 & 0.18 & 0.13 & 0.11 \\
SBG & $0.33\ldots$  & 8.61 & 0.55 & 0.38 & 0.11 \\
LMC & $0.33\ldots$ & 14.26 & 0.34 & 0.22 & 0.16 \\
MW & $0.33\ldots$ & 21.30 & 0.32 & 0.23 & 0.22 \\
GREY & $0.33\ldots$ & 24.80 & 6.18 & 7.10 & 0.15 \\
\hline
&\multicolumn{4}{l}{$L_{\nu} \propto \nu^{5/7}$} \\
\hline
SMC & $0.71\ldots$ & 5.99 & 0.28 & 0.15 & 0.11 \\
SBG & $0.71\ldots$ & 7.88 & 0.86 & 0.37 & 0.11  \\
LMC & $0.71\ldots$ & 24.30 & 0.52 & 0.23 & 0.21  \\
MW & $0.71\ldots$ & 40.44 & 0.48 & 0.27 & 0.29  \\
GREY & $0.71\ldots$ & 39.62 & 11.63 & 11.61 & 0.18  \\
\hline
&\multicolumn{4}{l}{Best-fit $\alpha$} \\
\hline
SMC & $0.71\pm0.02$ & 5.99 & 0.28 & 0.15 & 0.11 \\
SBG & $1.39\pm0.04$ & 7.37 & 1.43 & 0.39 & 0.12  \\
LMC & $-0.03\pm0.01$ & 11.19 & 0.18 & 0.21 & 0.15  \\
MW & $-0.13\pm0.01$ & 13.19 & 0.13 & 0.21 & 0.18  \\
GREY & $-0.39\pm0.02$  & 13.80 & -0.67 & 4.47 & 0.14  \\
\hline
\end{tabular*}
\caption{Summary of best-fit $\chi^2_{N}$ for the total sample of 9\,242 sources, reduced by the number of degrees of freedom ($N= 3\times9\,242$), including corresponding median $E(B-V)$ and $\sigma(g-i)$ for each of the five assumed dust laws, in the case where $L_{\nu} \propto \nu^{1/3}$, $\propto \nu^{5/7}$, and the best-fit spectral slope $\alpha$ whose uncertainty is boosted by $\sqrt{\chi^{2}/{N}}$.}
\label{tab:bestfit_params}
\end{table}

From the left panel of Fig.~\ref{fig:bestfit_chisq}, showing
the $\chi^2$ landscape as a function of $\alpha$,
we find that the graphite-heavy dust laws for \textsc{LMC} and \textsc{MW} are strongly disfavoured. Their minimum $\chi^{2}$ occurs for a red spectral slope ($\alpha < 0$) and under the assumption of $L_{\nu} \propto \nu^{1/3}$ they are disfavoured at high confidence in clear excess of 5$\sigma$.  The best-fit solution for the UV-flat \textsc{GREY} dust law occurs for an even redder
spectrum, $\alpha=-0.39$, and with an nonphysical median $E(B-V) =-0.67$~mag. \textsc{LMC}, \textsc{MW}, and \textsc{GREY} also have relatively high $\chi^{2}_{\rm min}$ values, corresponding to reduced $\chi^{2}/{N}$ values
11.19, 13.19, 13.80, respectively.

The \textsc{SMC} and \textsc{SBG} laws fare rather better, achieving lower best-fit $\chi^{2}_{\rm min}$ values of 5.99 and 7.37, 
respectively. The \textsc{SMC} dust law achieves its lowest overall $\chi^{2}_{\rm min}$ at $\alpha=0.71\pm0.02$ and is tightly constrained relative to the \textsc{SBG}, which achieves its relatively higher $\chi^{2}_{\rm min}$ 
at $\alpha=1.39\pm0.04$. While the \textsc{SMC} enjoys a smooth progression of median $E(B-V)$ values, the \textsc{SBG} requires even greater reddening for the same $\alpha$ despite turning over in $E(B-V)$ at the same $\alpha$. Thus, the aggregate sample measured with $\chi^{2}$ corroborates the aforementioned findings that the \textsc{SMC} provides the best-fit dust solution to describe the de-reddened disc SED of the sample considered. However, we also measure the total $\chi^2$ at each $\alpha$ by assigning each source a $\chi^2$ corresponding to the best-fit dust law, and recompute the best-fit $\alpha$ (called `BEST'). Unsurprisingly, this combined sample finds a minimum $\alpha=1.01\pm0.02$, mid way between the \text{SMC} and \text{SBG} laws which are the two best-fit dust laws for any given $\alpha$. However, this best-fit $\alpha$ is suffers from a high median $E(B-V)$ compared to the SMC law at $\alpha\sim1$, and may suffer from noisy measurements (as they are not weighted here), preferential $\chi^{2}$ arsing from unreasonably attenuated solutions, and other effects which make its interpretation non-trivial. 

The constraining power of $\chi^{2}$  in the left panel of Fig.~\ref{fig:bestfit_chisq} is distinctly superior to that of $\sigma(g-i)$ in the right panel. This makes sense, as $\chi^2$ utilizes all 5 bands compared to only 2 bands in $\sigma(g-i)$. Nevertheless, comparing these may increase confidence in the results and deepen our understanding of the models. Note that the order of the best-fit $\alpha$ values for different dust laws is the same for minima of  $\chi^{2}$ and minima of  $\sigma(g-i)$. The best-fit $\alpha$ values have a smaller range for minima of $\sigma(g-i)$.
The
\textsc{MW}, \textsc{LMC} and \textsc{GREY} dust laws cluster around $\alpha=0$, with relatively high $\sigma(g-i)$. The \textsc{SMC} and \textsc{SBG} are both close to $\alpha=0.7$, with \textsc{SBG} achieving a slightly lower $\sigma(g-i)$ than that for \textsc{SMC}.
The $E(B-V)$ values at a given $\alpha$ are generally similar between the estimators.

Owing to the large sample size of this investigation, the constraint on the best-fit slope $\alpha$ is remarkably tight, uncertain by of order 1\% for the $\chi^{2}$ badness-of-fit.
Both $\chi^2$ and $\sigma(g-i)$
prefer a power-law spectral index $\alpha$ significantly bluer than
 the canonical $L_{\nu} \propto \nu^{1/3}$ accretion disc spectrum. The $L_{\nu} \propto \nu^{1/3}$ power-law is statistically inconsistent with our results. 

The merit of this result is further explored in Fig.~\ref{fig:composite_spectrum2} where we perform the same spectral composition procedure as shown by Fig.~\ref{fig:composite_spectrum} but assume a $L_{\nu} \propto \nu^{5/7}$ while maintaining the assumption of an \textsc{SMC}-like power-law. By doing so, we find an even more consistent picture with the $L_{\nu} \propto \nu^{5/7}$ power-law, achieving a $\chi^{2}/{N}=9.00$. This is lower than that achieved with the expected $L_{\nu} \propto \nu^{1/3}$, suggesting that $\alpha=5/7$ is a more appropriate model. In addition, it is apparent that the bluest residuals have lessened, with the de-reddened disc spectrum now being consistent with a smooth $L_{\nu} \propto \nu^{5/7}$ power-law within a 68 percentile range for wavelengths bluer than $\mathrm{H}_{\alpha}$ and redder than the Lyman continuum break. We discuss the implications of this serendipitous finding in the following section.

\section{Discussion}\label{sec:discussion}

\subsection{Assumptions and caveats}

Advantageous properties of the SDSS data set analysed here are its unprecedented number of quasars and the long timespan over which the five-band $ugriz$ photometry has been obtained. For each quasar we leverage its variable nature to separate the variable disc component from its static host galaxy. Our decomposition method treats each observation as an independent measurement of the galaxy+disc flux at some time-dependent dimensionless brightness level $X(t)$, where $X=0$ is the mean level and $\Delta X=1$ is the rms of the lightcurve variations. This is illustrated in Fig.~\ref{fig:decomposition} where each flux measurement provides an independent constraint on the linear model, $F(\lambda,t)=B(\lambda) + A(\lambda)\, X(t)$.
Here the intercept $B(\lambda)$ is the mean galaxy+disc spectrum at $X=0$ and the slope $A(\lambda)$
is the rms spectrum of the disc variations.
Extrapolating the fit to fainter levels is assumed to
effectively turn off the variable disc light leaving 
just the galaxy spectrum at some minimum value of $X$.
This point is somewhat arbitrary, particularly when the variations are small so that the extrapolation is a long one.
In order to have a well defined decomposition,
we adopt the point at which the extrapolated flux is consistent with zero flux at 1$\sigma$, which we interpret as the limit below which the model is no longer physically meaningful.

Our estimates of $E(B-V)$ for five different extinction laws are computed for each source to quantify and compensate for extinction and reddening of the disc spectrum by dust along the line-of-sight to the accretion disc. This assumes that the observed disc spectrum is fainter and redder than the intrinsic disc spectrum due to line-of-sight reddening and extinction by dust, although the converse (requiring nonphysical negative attenuation) is an allowed solution.
For the intrinsic disc spectrum, we assume a power-law, $L_\nu \propto \nu^{\alpha}$. The estimate of $E(B-V)$ depends on the assumed power-law spectral index $\alpha$, a bluer slope requires a larger $E(B-V)$. 
Our power-law disc model neglects possible contributions of emission lines and bound-free continua. The variability of this approximation is supported by Fig.~\ref{fig:evol_panels}, which shows that the variable disc spectrum has weaker emission features than the mean spectrum. However, as shown by Fig.~\ref{fig:composite_spectrum}, the final de-reddened disc spectrum, assuming an \textsc{SMC}-like extinction and $L_{\nu} \propto \nu^{1/3}$, shows that some emission features remain. These are of course smoothed by the broad bandwidths of the $ugriz$ filters, leaving wide and weak rather than narrow and strong emission features in the residuals. Although visually the residual features corresponding to the $\alpha=1/3$ composite (Fig.~\ref{fig:composite_spectrum}) appear similar to those of the $\alpha=5/7$ (Fig.~\ref{fig:composite_spectrum2}), the latter achieves a significantly better fit, $\Delta\chi^{2}\sim20\,000$
or $\chi^{2}/N=11.75 \rightarrow 9.00$.
However, we caution that overfitting and unseen systematics may contribute to this effect.

Despite the straight-forward interpretation that a combination of SMC-like dust and a bluer spectral slope describes the variable accretion disc spectra of quasars, we note several caveats. First, for the \textsc{MW} and \textsc{LMC} laws the  $E(B-V)$ distribution of the SDSS quasars has an implausible redshift dependence caused by the strong rest-frame 2175\,\AA\, absorption feature moving across the center of a band. A ``beating'' pattern is observed where the estimates of $E(B-V)$ have a large scatter when the 2175\,\AA\, feature is not directly observed (see Appendix~\ref{sec:app1}). This highlights a shortcoming in the modeling of the dust when adopting the \textsc{MW} and \textsc{LMC} models. We considered addressing this by using a prior favouring models that make $E(B-V)$ a smoother function of redshift, but decided in the interest of simplicity to omit this complication in our modelling. 
Our model also places no limitation on the extent to which the variable component can be reddened, which may permit extreme reddening requiring extraordinary dust column densities. We note that, with the exception of the \textsc{GREY} law, there are no instances of problematically dusty attenuation estimates. 

\subsection{On attenuation laws}

The dust laws investigated in this work broadly fall into two categories. Either they are well-described by a smooth power-law-like curve (e.g., \textsc{SMC}, \textsc{SBG}), or a power-law-like curve with a strong graphite absorption feature at 2175\,\AA\,(\textsc{LMC}, \textsc{MW}). The outlying case is that of the Gaskell's dust law derived from a sample of AGN (\textsc{GREY}). Their differences are highlighted in Fig.~\ref{fig:dustlaws}. 

As shown for a specific case in Fig.~\ref{fig:dereddening}, the likelihood that a particular dust law is well-suited for a particular source is assessed here using $\chi^2$ to quantify the badness-of-fit and $E(B-V)$ to indicate cases where exceptional dust columns would be required. Similarly, to quantify the success in modelling the SDSS quasar sample as a whole, we employ two badness-of-fit metrics, $\chi^2$ and $\sigma(g-i)$,  along with the median $E(B-V)$.
Given that the information presented by $\sigma(g-i)$ is contained in $\chi^{2}$ and has generally less constraining power (see Fig.~\ref{fig:bestfit_chisq}), we adopt $\chi^{2}$ as the primary criterion, using 
$\sigma(g-i)$ and $E(B-V)$ for secondary considerations.

As presented in Section~\ref{sec:application}, assuming a $L_{\nu} \propto \nu^{1/3}$ power-law, the least likely dust-law is \textsc{GREY} which is statistically excluded at high confidence for the vast majority of the sources. However, 7\% of the sample finds a best-fit solution with the \textsc{GREY} extinction law, but with an exceptionally large median $E(B-V)=6.86$~mag for this sub-sample. Further, we find evidence to exclude the graphite absorption laws of the \textsc{LMC} and \textsc{MW} as a general best-fit, finding the best-fit solution for only 17\% and 16\% of the total sample, respectively. 

We find the greatest success with the smooth power-law extinction laws, \textsc{SMC} and \textsc{SBG}. As measured by $\chi^{2}$ in Fig.~\ref{fig:bestfit_chisq}, the \textsc{SMC} provides the best fit to the sample as a whole and is consistent with a best-fit $\alpha\sim0.7$ from both the $\chi^2$ and $\sigma(g-i)$ estimators. Assuming an  $L_{\nu} \propto \nu^{1/3}$ power-law, Fig.~\ref{fig:dustfractions} shows that the \textsc{SMC} provides the best-fit solution for 43\% of the sample while the \textsc{SBG} provides 17\%. Interestingly, although constrained with less information, the $\sigma(g-i)$ estimator finds that the \textsc{SBG} provides a solution similar to the \text{SMC} consistent with a $L_{\nu} \propto \nu^{5/7}$ power-law. Thus, while we cannot exclude the \textsc{SBG} law outright, we nonetheless find the most likely dust law for this sample of quasars is the \textsc{SMC}. This result holds also in the case of an assumed $L_{\nu} \propto \nu^{5/7}$ power-law.

Considering now the derived composite SEDs shown in Figures~\ref{fig:composite_spectrum} and \ref{fig:composite_spectrum2}, under the assumption of an \textsc{SMC}-like dust law derived assuming either $L_{\nu} \propto \nu^{1/3}$ or $\propto \nu^{5/7}$, there are no discernible features consistent with strong Balmer absorption or emission at the 10\% level. However, given that the emission is smeared across the broad-band filter, only the highest equivalent width lines could be detected. Continuum consistent with thermal emission from an optically-thick accretion disc is clear \citep[i.e., the `Big Blue Bump';][]{Malkan1982}. 

The Southern Sample data set contains several sources observed at $3<z<6$. At these redshifts, the rest-frame $u$ band intersects the expected rest-frame UV turnover of the accretion disc spectrum at $\sim1000\,\AA$. Although this subset constitutes only a small fraction of the total sample, the effect of the turnover is evident from Fig.~\ref{fig:composite_spectrum}. The handful of these sources approaching $z\sim6$ may also be affected by the neutral intergalactic medium which may be contributing to this observed turnover with resonant absorption by hydrogen gas clouds along the line-of-sight (i.e., the well-known Lyman Forest). Regardless of the physical mechanism driving this highly significant turnover, it is not accounted for in the continuous power-law form assumed for the disc spectrum. Consequently, the continuous power-law model is not appropriate for the bluest bands for $z>3$ objects, whose $E(B-V)$ may be overestimated due to the turnover acting as an extreme reddening of the $u$ band.

\subsection{Best-fit power law exponent}

Initially we assumed the canonical $L_{\nu} \propto \nu^{1/3}$ power-law before expanding to a range of power-law exponents to determine the most suitable power-law index and extinction law combination as assessed by the badness-of-fit $\chi^{2}$ statistic. The result is shown in Fig.~\ref{fig:bestfit_chisq}. The \textsc{SMC} law shows remarkable agreement from both $\chi^2$ and $\sigma(g-i)$ estimators finding a best-fit blue slope $\alpha\sim0.7$ in both cases. For the latter estimator, \textsc{SBG} finds its minimum also at $\alpha\sim0.7$. Taking $\chi^2$ to be the more robust and more precise estimator, the corresponding $\alpha$ for the best-fit \textsc{SMC} is 0.71$\pm$0.02. This is highly inconsistent with $L_{\nu} \propto \nu^{1/3}$ predicted for a geometrically thin steady-state disc, within theses strict uncertainties.

Given this unexpected result, we verified the robustness of the de-reddening procedure by measuring 9000 simulated disc SEDs with $\alpha=1/3$ perturbed with random noise corresponding to that of the observed photometry. We successfully recovered a best-fit power-law index of $1/3$, confirming that the procedure is unbiased and that the result is indeed genuine.

This result contrasts with work from \citet{Kokubo2014}, who used the same SDSS data set and employed a 'flux-flux' deomposition method most similar to ours in order to explore the variations in the photometric lightcurves. They found that the composite spectrum \textit{before} decomposition features a red, $\alpha=-0.5$ slope relative to the much bluer $\alpha=1/3$ `difference' composite spectrum, in agreement with \citep{ShakuraSunyaev1973}. However, \citeauthor{Kokubo2014} caution that their method does not attempt to estimate or account for non-variable components of the host galaxy. Despite this work being the closest analog to the present study available in the literature, the still different methodologies and assumptions make a concrete comparison of derived spectral slopes hazardous. It is possible though that the bluer slope derived in the present work is found because we removed the static host galaxy component, which would otherwise produce a redder best-fit spectral slope.

Entirely serendipitously, we find our resulting $\alpha=0.71\pm0.02$ to be statistically consistent with the recent theoretical framework proposed by \citet{Mummery20_tde} who predict an mid-range spectral slope $\alpha=5/7\simeq0.71$. We caution however that this 1) assumes that all quasars have SMC-like dust and 2) is sensitive to the tail end of the $\chi^2$ distribution used to compute the total $\chi^2$ from Fig~\ref{fig:bestfit_chisq}. At face value, however, the model proposed by \citeauthor{Mummery20_tde}{} cannot be ruled out by our results.

This finding is intriguing, as it hints at additional accretion physics not considered in many previous studies. \citeauthor{Mummery20_tde} propose a fully-relativistic framework of accretion discs, finding a temperature structure driven by energy liberated by viscous and magnetically-driven torques arising from a spinning black hole, and which appear at the innermost stable circular orbit.  This non-vanishing stress term provides the necessary energy, in addition to the typical gravitational energy of the momentum transfer of the disc, to steepen the implied temperature structure. The result is a spectral slope which is a factor of $\sim2\times$ larger (i.e. bluer) than in traditional $L_{\nu} \propto \nu^{1/3}$ steady-state accretion models.

\subsection{ Comparison with Intensive Disc Reverberation Mapping}

An independent method being used to probe accretion disc temperature profiles is to
obtain intensive (sub-day) monitoring and then measure inter-band time delays
\citep{2007MNRAS.380..669C, 2019ApJ...870..123E}.
 This intensive disc reverberation mapping method (IDRM) assumes light travel time delays $\tau \approx R/c$,
and blackbody emission peaking near $\lambda\approx( h\,c/k\,T)$.
A reverberating disc with $T\propto R^{-b}$ gives time delays $\tau \propto \lambda^{1/b}$, and disc spectra 
$L_\nu \propto \nu^{(2b-3)/b}$.
The standard disc model with $b=3/4$ predicts $\tau \propto \lambda^{4/3}$
and $L_\nu \propto \nu^{1/3}$, while a steeper temperature profile with $b=7/8$
predicts $\tau\propto\lambda^{8/7}$
and $L_\nu\propto\nu^{5/7}$.

The IDRM results to date are typically described as consistent with the standard disc prediction, $b=3/4$, but with uncertainties large enough to admit $b=1$.
The most accurate IDRM results to date,
from the Space Telescope and Optical
Reverberation Mapping campaign monitoring of NGC~5548 in 2014, give a best-fit
power-law slope $b=1.03\pm0.12$
from cross-correlation lags \citep{2016ApJ...821...56F}
or $b=0.99\pm0.03$ from detailed fitting
of a reverberating disc model to the lightcurves \citep{2017ApJ...835...65S}.
This corresponds to a steeper temperature profile, closer to $b=7/8$ ($\alpha=5/7)$ than to $3/4$~($1/3$).

A caveat, however, is that the disc sizes 
inferred from IDRM are typically larger than expected, by a factor of $\sim3$, and an excess lag in the Balmer continuum region suggests that bound-free emission from the (larger) broad emission-line region (BLR) may be
contributing significantly to the  cross-correlation lags \citep{Korista2001, Lawther2018}.
Perhaps the clearest example is the lag spectrum from HST monitoring of NGC~4593 \citep{2018ApJ...857...53C}.
Work is underway to understand how best to disentangle the
disc and BLR contributions to the measured lags.

\section{Summary \& Conclusions}
\label{sec:conclusions}

In this work, we have separated the variable accretion disc light from their static host galaxies using broad-band photometric lightcurves, de-reddened their SEDs to account for dust, and leveraged the results to test accretion disc physics.

\begin{itemize}
  \setlength\itemsep{1em}
  \item We developed a method for decomposing quasar lightcurves by separating the contribution of the variable light from the static, background emission. Each disc SED was then de-reddened to provide a dust-free estimation of the underlying accretion disc light.
  \item Of the five dust laws examined in this work, we find that the featureless laws of the Small Magellanic Clouds \citep[SMC;][]{Gordon2003} and starburst galaxies \citep[\textsc{SBG};][]{Calzetti2000} are the most reasonable attenuation models as measured by their $\chi^2/N$, with the SMC being slightly preferable as it typically requires less attenuation than the \textsc{SBG}.
  \item Assuming an SMC-like dust attenuation, the best-fit spectral slope $\alpha$ is found to be inconsistent with a standard $L_{\nu} \propto \nu^{1/3}$ corresponding the steady-state accretion model  \citep{ShakuraSunyaev1973}. Instead, we find significant evidence for a $L_{\nu} \propto \nu^{5/7}$ accretion slope based on our best-fit $\alpha = 0.71\pm0.02$, in agreement with the proposed disc model of \citet{Mummery20_tde}. 
\end{itemize}

If it is indeed the case that the model of \citeauthor{Mummery20_tde} better reflects the reality of accretion physics than previous models, then these observational findings challenge commonly made assumptions about the thermal structure of quasar accretion dics. Moreover, they have implications for deriving key properties of black holes and their accretion discs, including the Eddington luminosity and black hole mass. Future work is needed to confirm this model, requiring independent observations and continued monitoring of key sources in a way which these results can be confidently replicated. 

Finally, we note that the methodology developed here should be useful for analysis of quasar variability data from the LSST, and could also be applied to data from multi-object spectroscopic monitoring surveys such as SDSS-RM.

\section*{Acknowledgements}
The authors would like to thank 
Chelsea Macleod,
Marianne Vestergaard,
and
Michael Goad
for helpful discussions.
We would also like to thank the organisers of the 2019 Quasars in Crisis meeting for facilitating so many useful conversations. Lastly we would like to thank the anonymous referee for their generous insight and helpful comments which improved this work.

The Cosmic Dawn Center (DAWN) is funded by the Danish National Research Foundation under grant No. 140. J.W. acknowledges support from the European Research Council (ERC) Consolidator Grant funding scheme (project ConTExt, grant No. 648179), and from the University of St Andrews Undergraduate Research Assistant Scheme. K.H. acknowledges support from STFC grant ST/R000824/1.

Funding for the SDSS and SDSS-II has been
provided by the Alfred P. Sloan Foundation,
the Participating Institutions, the National Science Foundation, the U.S. Department of Energy,
the National Aeronautics and Space Administration, the Japanese Monbukagakusho, the Max
Planck Society, and the Higher Education Funding Council for England. The SDSS Web Site is
\url{http://www.sdss.org/}.
The SDSS is managed by the Astrophysical
Research Consortium for the Participating Institutions. The Participating Institutions are the
American Museum of Natural History, Astrophysical Institute Potsdam, University of Basel,
University of Cambridge, Case Western Reserve
University, University of Chicago, Drexel University, Fermilab, the Institute for Advanced
Study, the Japan Participation Group, Johns
Hopkins University, the Joint Institute for Nuclear Astrophysics, the Kavli Institute for Particle Astrophysics and Cosmology, the Korean
Scientist Group, the Chinese Academy of Sciences (LAMOST), Los Alamos National Laboratory, the Max-Planck-Institute for Astronomy (MPIA), the Max-Planck-Institute for Astrophysics (MPA), New Mexico State University,
Ohio State University, University of Pittsburgh,
University of Portsmouth, Princeton University,
the United States Naval Observatory, and the
University of Washington.

\section*{Data Availability}
The data underlying this article were accessed from \href{http://faculty.washington.edu/ivezic/macleod/qso_dr7/index.html}{Southern Survey of Stripe 82 Quasars} as part of the Sloan Digital Sky Survey. The derived data generated in this research will be shared on reasonable request to the corresponding author.




\bibliographystyle{mnras}
\bibliography{bibliography} 




\appendix

\section{Physical Interpretation of Attenuation Estimates}

\begin{figure*}
	\resizebox{\hsize}{!}{\includegraphics[width=\columnwidth]{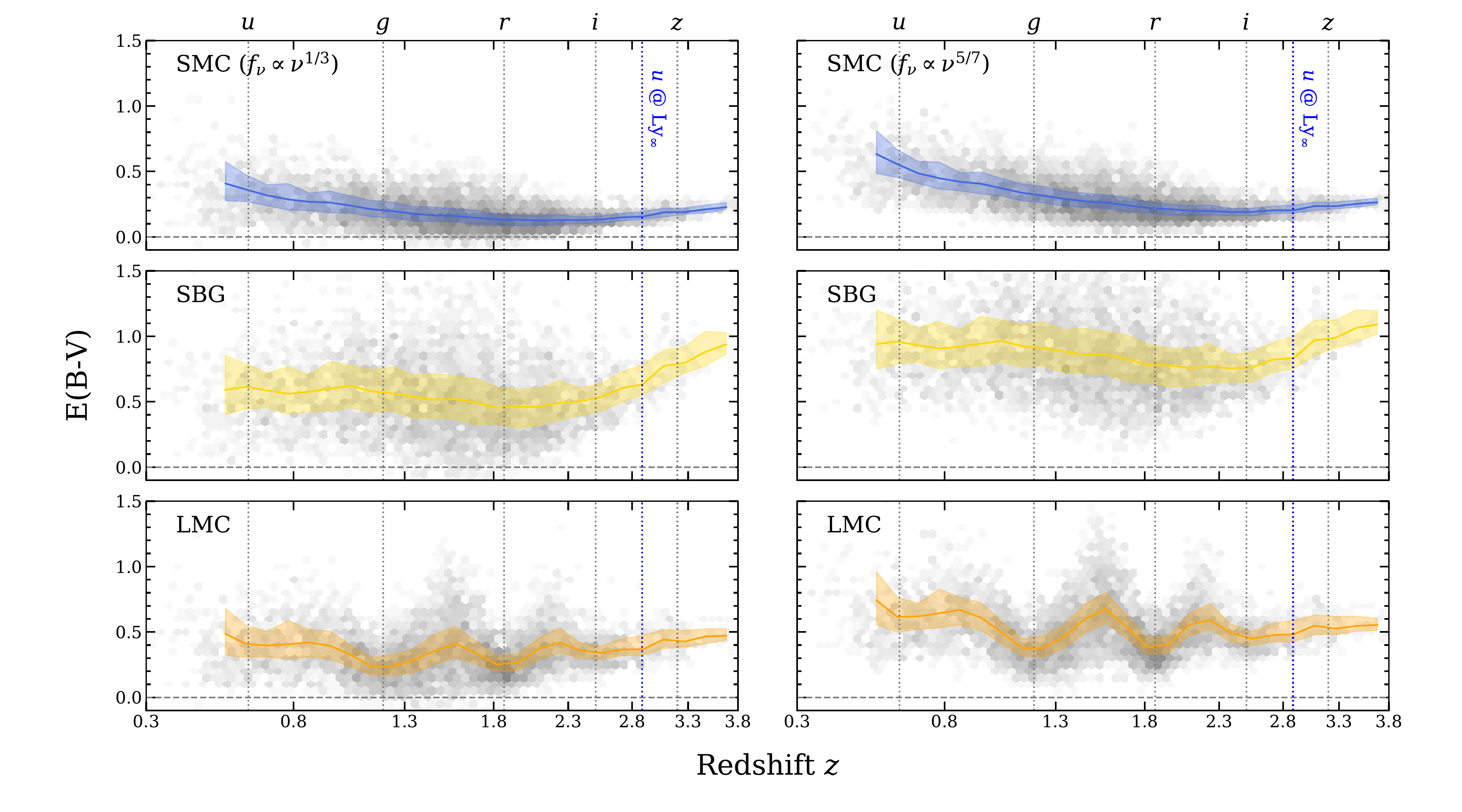}}
	\caption{Binned median estimates of $E(B-V)$ over a range of redshift, assuming a $\mathrm{F}_{\nu}\,\sim\,\nu^{1/3}$ (left) or a $\mathrm{F}_{\nu}\,\sim\,\nu^{5/7}$ (right), for the two best performing extinction laws (\textsc{SMC}, \textsc{SBG}) as well as the \textsc{LMC} which features significant absorption at 2175\,\AA. The redshift at which the 2175\,\AA\, feature is centered in each band is shown by the grey dotted lines. The coloured envelopes contain 68\% of sources in each bin.}
	\label{fig:dust_evol}
\end{figure*}

As discussed briefly in Section~\ref{sec:discussion}, we do not place any priors on the allowed ranges of $E(B-V)$ parameter in our dereddening procedure. As a consequence of the prominent 2175\,\AA\, feature in the \textsc{LMC} and \textsc{MW} dust laws, estimates of $E(B-V)$ are more similar for sources where the bump is directly constrained by one of the five bands, which is shown to undulate with redshift in Fig.~\ref{fig:dust_evol} for the \textsc{LMC}. This undulation is not seen, however, in the smooth dust laws of the \textsc{SMC} and \textsc{SBG}. We also find $E(B-V)$ flares up for sources at $z>3$ where the $u$-band falls blueward of the Lyman continuum break. This discontinuity is not included in our continuous power-law model, and so mimics an extreme reddening. We do not interprete either feature as a genuine physical phenomena of accretion discs, but merely a limitation of our model. The consequences of this are described in Section~\ref{sec:discussion}.

\label{sec:app1}



\bsp	
\label{lastpage}
\end{document}